\pdfoutput=1
\documentclass[12pt,a4paper]{article}
\usepackage[usenames, dvipsnames]{color}
\usepackage{graphics,graphicx,psfrag}
\usepackage{amsmath,amssymb}
\usepackage{bm}	
\usepackage{verbatim}
\usepackage{jheppub}
\usepackage{dsfont}
\usepackage{multirow}
\usepackage{subfigure}
\usepackage{slashed}
\usepackage{ulem}
\usepackage{color}
\usepackage{ulem}

\def\issue(#1,#2,#3){{\bf #1}, #2 (#3)}

\catcode`\@=11
\def\lsim{\mathrel{\mathpalette\@versim<}}
\def\gsim{\mathrel{\mathpalette\@versim>}}
\def\@versim#1#2{\vcenter{\offinterlineskip
\ialign{$\m@th#1\hfil##\hfil$\crcr#2\crcr\sim\crcr } }}
\catcode`\@=12

\catcode`@=12

\newcommand{\nn}{\nonumber\\}
\newcommand{\bb}{\begin{bmatrix}}
\newcommand{\eb}{\end{bmatrix}}

\newcommand{\Dslash}{D\!\!\!\!/\,}
\newcommand{\order}{ {\cal O} }
\newcommand{\be}{\begin{equation}}
\newcommand{\ee}{\end{equation}}

\def\issue(#1,#2,#3){{\bf #1}, #2 (#3)}

\hypersetup{
   colorlinks=true,       
   linkcolor=blue,        
   citecolor=red,         
   filecolor=magenta,      
   }

\preprint{\begin{flushright}UCI-HEP-TR-2019-02 \\ MSUHEP-19-007\end{flushright}}

\title{Direct Detection and LHC constraints on a $t$-Channel Simplified Model 
of Majorana Dark Matter at One Loop}

\author[a]{Kirtimaan A. Mohan,}
\author[a,b]{Dipan Sengupta,} 
\author[c]{Tim M.P Tait,}
\author[a]{Bin Yan,}
\author[a]{C.--P. Yuan}

\affiliation[a]{Department of Physics and Astronomy, Michigan State University, 567 Wilson Road, East Lansing , U.S.A}
\affiliation[b]{Department of Physics and Astronomy, University of California, San Diego, 9500 Gilman Drive, La Jolla, U.S.A}
\affiliation[c]{Department of Physics and Astronomy, University of California, Irvine, CA 92697-4575 USA}

\abstract{
An interesting class of models posits that the dark matter is a Majorana fermion which interacts with a quark together
with a colored scalar mediator.   Such a theory can be tested in  direct detection experiments,
 through dark matter scattering with  heavy nuclei, and at the LHC,  via 
 jets and missing energy signatures.
   Motivated by the fact that such theories have spin-independent interactions that vanish
at tree level, we examine them at one loop (along with RGE improvement to resum large logs), 
and find that despite its occurrence at a higher order of perturbation theory,
the spin-independent scattering searches typically impose the strongest constraints on the model parameter
space.  We further analyze the corresponding LHC constraints  at one loop and find that it is important to take them into account
when interpreting the implications of searches for jets plus missing momentum on this class of models, thus providing 
the corresponding complementary information for this class of models.  }

\emailAdd{kamohan@pa.msu.edu,disengupta@physics.ucsd.edu,ttait@uci.edu,\\ yanbin1@msu.edu,yuan@pa.msu.edu}

\begin{document}

\maketitle

\section{Introduction}
\label{sec:intro}

Observations from cosmology and large scale structures indicate that the Universe is filled with a non-relativistic species of
particle that so far appears to be completely transparent to photons of all wavelengths.  The properties
of this dark matter (DM) appear to be inconsistent with any ingredient known to the Standard Model (SM) of particle physics,
and thus represents a glimpse of physics beyond it.  
If the dark matter has appreciable interaction with the SM fields, its abundance in the Universe today can be understood 
as the result of a thermal freeze-out process. Based on this hope,
there is a major effort currently underway to detect it
through its annihilation products, scattering with ordinary matter (such as heavy nuclei), or by producing it directly at high energy colliders.

Assembling a complete picture of what the dark matter can (or cannot) be requires us to collate information from all sources.
Understanding how to map one class of search into another one requires a theoretical framework. Within the thermal freeze-out paradigm,  there are a variety of
possible theories of dark matter, ranging in completeness from effective field theories to simplified models
to UV complete theories such as supersymmetry.  Recently, simplified models have emerged as a robust mechanism to
contrast various particle searches, as they balance a reasonable simple theoretical framework with a finite number
of parameters against enough detail to encapsulate a theoretically complete description valid up to the energies
probed a the Large Hadron Collider (LHC) ~\cite{Abdallah:2015ter}.

There are a variety of simplified models employed to interpret LHC searches, largely classified by the nature of the mediator which
communicates between the SM and the dark sectors.   
Much of previous work has considered ``$s$-channel models"
in which the mediator is a dark force carrier, a neutral
boson which has interactions both with a pair of dark matter particles and with a pair of Standard Model particles
\cite{Petriello:2008pu,Bai:2010hh,Boveia:2016mrp,Abe:2018bpo}.   While interesting parameter space remains to be
explored, such constructions are generically constrained by searches for visible decays of the
mediator \cite{Dreiner:2013vla,Albert:2017onk}.  A different, and equally compelling class of models
contains colored mediator particles, which can either
interact directly with a quark and a dark matter 
particle~\cite{Chang:2013oia,An:2013xka,Bai:2013iqa,DiFranzo:2013vra,Bell:2015sza,Ibarra:2015fqa,Garny:2015wea,Ko:2016zxg,Mandal:2018czf,Biondini:2018ovz},
or act as a bridge at loop level between a pair of dark matter particles and a pair of gluons \cite{Godbole:2015gma,Bai:2015swa,Godbole:2016mzr}.  
Such colored mediators are in principle
accessible at the LHC, leading to signatures of missing momentum accompanied by energetic jets of hadrons,
and are necessary ingredients in UV complete models of physics beyond the Standard Model such as
supersymmetry or Little Higgs models.
At the same time, in contrast to the $s$-channel models, there are no purely SM searches to restrict the 
viable parameter space.  They are thus important to understand the relative importance of jets plus missing momentum
searches to probe dark matter models.  Simultaneously, direct detection experiments also provide strong 
constraints on the parameter space of these models, thus providing complementary information. Current and next 
generation direct detection experiments will probe a large part of the Weakly Interacting Massive Particle
(WIMP) regime of the dark matter landscape,  thus providing a better understanding of the theoretical space of models.

In order to correctly parse the implications for experimental searches on the parameter space of dark matter models, accurate
theoretical predictions are required.
This is particularly important in the case where the dark matter is a Majorana fermion
and the mediator is a colored scalar particle.  In that case, the tree level contribution to the 
spin-independent scattering with nuclei
vanishes, leaving much weaker constraints from spin-dependent searches~\cite{DiFranzo:2013vra}.  But contributions
to the spin-independent scattering rate still occur at one loop level, and as we shall see below, represent the  dominant
constraints for wide regions of parameter space.
In this article, we extend our understanding of this simplified model to the next-to-leading-order:
\begin{itemize}
\item We compute the one loop (leading non-vanishing) order contribution to the spin-independent scattering operator,
and perform renormalization group evolution (RGE) 
from high scales of order the mediator mass down to the low energy scales relevant for
dark matter scattering with a heavy nucleus.
\item We compute the LHC production cross sections to next-to-leading order (NLO) in $\alpha_S$, and recast the existing LHC searches into the
simplified model parameter space.
\end{itemize}
Our results demonstrate that these refinements significantly alter the remaining viable parameter space in light of the null
searches for dark matter scattering, and also make important changes to the impact and prospects of searches at the LHC. 
In particular, we find that the picture based on the leading order scattering changes by roughly an order of magnitude
when next-to-leading order contributions are included.  The impact on limits from the LHC is less dramatic, but nonetheless DM production rates can change
by as much as $\sim 50\%$, leading to very significant impact on the allowed parameter space.

The rest of this paper is organized as follows. In Sec. \ref{sec:model} we describe the simplified model and the assumptions
concerning the parameter space.  In Sec. \ref{sec:blo}, we discuss the scattering with heavy nuclei, including 
the NLO contributions and RGE evolution of the Wilson coefficients. 
In Sec \ref{sec:constraints}, we assess the complementary collider constraints originating 
from LHC searches. In Sec, \ref{sec:combination}, we provide a summary of all constraints, as well as compute 
the velocity averaged annihilation cross section in order to assess whether this class of models can provide 
the correct thermal relic. 
 Finally we present our concluding remarks  in Sec. \ref{sec:outlook}. 

\section{Simplified Model and Parameters}
\label{sec:model}

In this section, we briefly review the simplified model,  more  details of which can be found in~\cite{DiFranzo:2013vra}.
The simplified model contains a SM singlet fermionic dark matter candidate ($\chi$), whose
kinetic terms are described by the Lagrangian
\begin{equation}
\mathcal{L}_{\chi}= \frac{1}{2} \left( i\bar{\chi}\slashed{\partial}\chi - M_{\chi}\bar{\chi}\chi\right)\ .
\end{equation}
While $\chi$ can be either Dirac or Majorana,  we specialize to the Majorana case where large
corrections are expected to the cross section for scattering with nuclei.
There are also a set of scalar mediator particles, which,
to interact with the dark matter and a SM quark,
must be color triplets transforming under the electroweak symmetry as (using notation $( SU(3), SU(2) )_Y$):
\begin{eqnarray}
(3,1)_{2/3}, ~~~~~ (3,1)_{-1/3}, ~~~~~ (3,2)_{-1/6}.
\end{eqnarray}
These three choices correspond to what we will refer to as a $u_R$ model (with mediators labeled as $\tilde{u}$),
a $d_R$ model (with mediators $\tilde{d}$), and a $q_L$ model (with mediators $\tilde{Q}$), respectively.
Motivated by the assumption of minimal flavor violation (MFV)~\cite{DAmbrosio:2002vsn}, 
we assign the mediators to flavor triplets with equal masses and couplings.
Thus the mediator and its dynamics can be described by the corresponding choice of Lagrangian:
\begin{equation}
\mathcal{L} _{u_R}= 
\sum_{u} \left[ (D_{\mu}\tilde{u})^*(D^{\mu}\tilde{u}) - M_{\tilde{u}}^2 ~ \tilde{u}^* \tilde{u}
+  g_{DM}~\tilde{u}^{*} ~\bar{\chi}P_{R} u + g_{DM}^*~\tilde{u} ~\bar{u}P_{L}\chi \right]\ ,
\label{eq:ur}
\end{equation}
\begin{equation}
\mathcal{L} _{d_R}= 
\sum_{d} \left[ (D_{\mu}\tilde{d})^*(D^{\mu}\tilde{d}) - M_{\tilde{d}}^2 ~\tilde{d}^* \tilde{d}
+ g_{DM}~\tilde{d}^{*} ~\bar{\chi}P_{R} d + g_{DM}^*~\tilde{d} ~\bar{d}P_{L}\chi \right]\ ,
\label{eq:dr}
\end{equation}
\begin{equation}
\mathcal{L} _{q_L}=
\sum_{q} \left[ (D_{\mu}\tilde{q})^*(D^{\mu}\tilde{q}) - M_{\tilde{q}}^2~ \tilde{q}^* \tilde{q}
+ g_{DM}~\tilde{q}^{*} ~\bar{\chi}P_{L} q + g_{DM}^*~\tilde{q}~\bar{q}P_{R}\chi)\right]\ ,
\label{eq:ql}
\end{equation}
where the covariant derivative
$D_{\mu}= \left(\partial_{\mu} - i g_sG^{a}_{\mu}T^{a} +~{\rm Electroweak~terms} \right)$,
describes the mediator couplings to the SM gauge bosons.
Here the sums are over quark and mediator flavors where $u = \left\{u,d,s\right\}$ quarks, $\tilde u = \left\{\tilde u,\tilde d,\tilde s\right\}$ mediators,  $d = \left\{d,s,b\right\}$~quarks, $\tilde d = \left\{\tilde d,\tilde s,\tilde b\right\}$ mediators, $q=\left\{u,d,s,c,b,t  \right\}$~quarks and $\tilde q=\left\{\tilde u,\tilde d,\tilde s,\tilde c,\tilde b,\tilde t  \right\}$ mediators.
In order to have a dark matter candidate which is uncolored or charged, we restrict ourselves to the parameter space in which all of
the mediator masses are larger than $M_\chi$.

Generation-dependent masses and couplings 
that are higher order in the Yukawa couplings can be generated consistently with MFV and can be described (for example, for the $\tilde{u}_R$ model) by terms such as:
\begin{equation}
\mathcal{L}_{FV}= \left(
\delta g_{DM}~ \tilde{u}^{*} Y^{u\dagger} Y^{u} \bar{\chi}P_{R}u + h.c.) 
+ \delta m^2 ~ \tilde{u}^{*} Y^{u\dagger} Y^{u} \tilde{u} + \mathcal{O}(Y^4).
\right) \ ,
\end{equation} 
where $Y^u$ is the SM Yukawa matrix. 
For simplicity and to avoid potential strong constraints from the null results of searches
for flavor and CP-violation, we take $\delta g_{DM} = \delta m^2 = 0$
and choose $g_{DM}$ to be real, from here onward.

\section{Scattering with Heavy Nuclei}
\label{sec:blo}

In the non-relativistic limit, dark matter scattering with a nucleus is described by a spin-independent (SI) term,
which at low momentum transfer resolves the entire nucleus coherently leading to a cross section enhanced by the
squared number of scattering centers (nucleons); and a spin-dependent (SD) term, which couples to the nucleon spin
and typically enjoys no coherent enhancement for large nuclei.

\begin{figure}
	\centering
\includegraphics[width=0.45\textwidth]{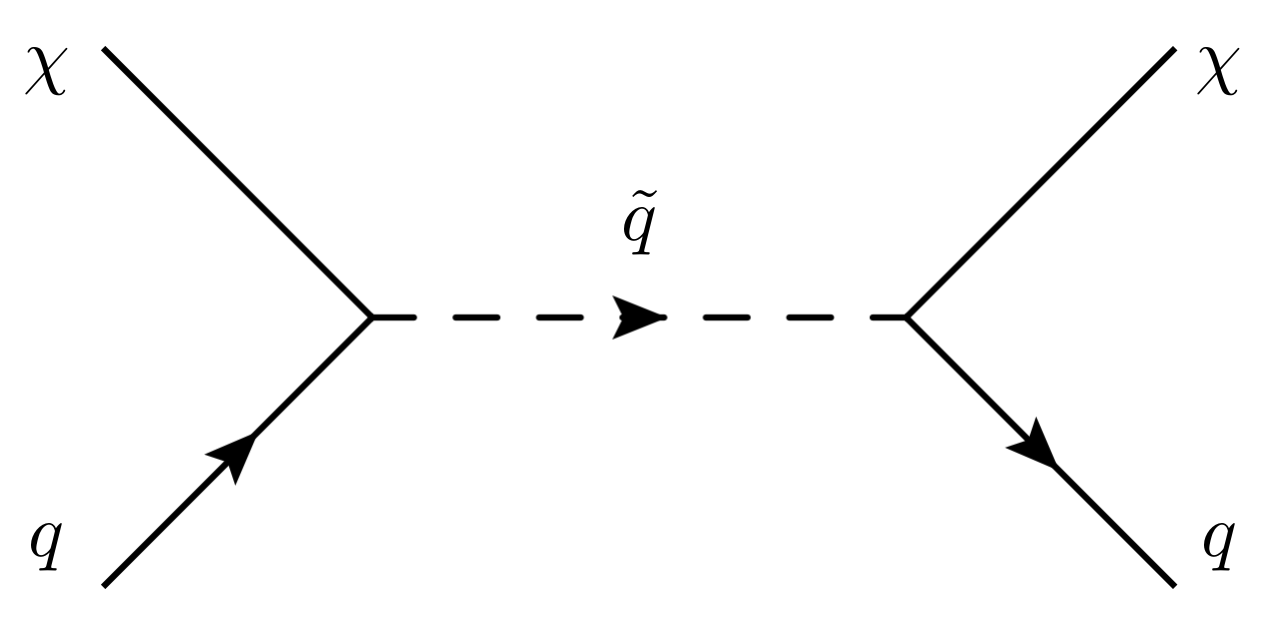}
\caption{Representative Feynman diagram for scattering between DM and quarks. \label{fig:feynqLO} }
\end{figure}

We begin by reviewing some of the results of Reference~\cite{DiFranzo:2013vra}, 
which represents the baseline upon which our improvements build. 
At leading order (LO), the dark matter interacts with a generic quark $q$ via tree level exchange of its corresponding mediator $\tilde{q}$,
as shown in a representative Feynman diagram in Figure~\ref{fig:feynqLO}.
The partonic matrix element
for interactions between fermionic (Dirac or Majorana) dark matter
and up-type quarks mediated by $\tilde{u}$ takes the form,
\begin{gather}
\mathcal{M} = (-ig_{\!_{DM}})^2 (\bar{\chi}P_Ru) \frac{i}{p^2-M_{\tilde{u}}^2} (\bar{u}P_L\chi) \nonumber\\
\approx (-ig_{\!_{DM}})^2 (\bar{\chi}P_Ru) \frac{-i}{M^2_{\tilde{u}}-m^2_{\chi}} (\bar{u}P_L\chi) + \mathcal{O}\left( \left[\frac{1}{M^2_{\tilde{u}}-m^2_{\chi}}\right]^2\right) \nonumber\\
\begin{split}
=\frac{ig_{\!_{DM}}^2}{M^2_{\tilde{u}}-m^2_{\chi}} \frac{1}{8} [(\bar{\chi}\gamma^{\mu}\chi)(\bar{u}\gamma_{\mu}u) - (\bar{\chi}\gamma^{\mu}\gamma^5\chi)(\bar{u}\gamma_{\mu}\gamma_5u) \nonumber\\ + (\bar{\chi}\gamma^{\mu}\gamma^5\chi)(\bar{u}\gamma_{\mu}u) - (\bar{\chi}\gamma^{\mu}\chi)(\bar{u}\gamma_{\mu}\gamma_5u)]
\end{split}\\
\approx \frac{ig_{\!_{DM}}^2}{M^2_{\tilde{u}}-m^2_{\chi}} \frac{1}{8} [(\bar{\chi}\gamma^{\mu}\chi)(\bar{u}\gamma_{\mu}u) - (\bar{\chi}\gamma^{\mu}\gamma^5\chi)(\bar{u}\gamma_{\mu}\gamma_5u)]
\label{eq:MLO}
\end{gather}
where, in the second line, the propagator is expanded in the low momentum limit and only leading terms are kept. 
As discussed in Section~\ref{sec:WilsonCoefficients} and Appendix~\ref{app:WilsonCoefficients},
higher order terms (which were dropped in Reference~\cite{DiFranzo:2013vra}) turn out to be important.  
In the last line of Equation~\ref{eq:MLO}, we have dropped terms which are negligible in the non-relativistic limit.
Furthermore, we have dropped the quark mass from the expressions above to simplify them.
Majorana fermions are treated using the technology of Refs.~\cite{Denner:1992me,Denner:1992vza}.
Analogous results as above hold for $d_R$ and $q_L$ quarks mediated by $\tilde{d}$ and $\tilde{q}$, respectively.
The terms in the last line result in spin independent and spin dependent scattering, respectively. 
However, since a Majorana fermion has a vanishing vector bilinear $(\bar{\chi}\gamma^{\mu}\chi = 0)$, only the SD terms are non-zero at this order\footnote{
It is worth noting that this feature is a consequence of having a single type of mediator.  In theories with both $\tilde{Q}$ and either $\tilde{u}$ or $\tilde{d}$ type
mediators, there may be renormalizable interactions involving both mediators and a Higgs boson, which would open up the possibility for tree level spin-independent scattering.}.
In order to assess the rate of SI scattering, it is necessary to go beyond the simple leading order calculation.

Following the notation of Refs.~\cite{Hisano:2010ct} and~\cite{Hill:2014yka} we write down the lagrangian for the effective field theory describing
SI interactions with quarks and gluons,
\begin{eqnarray}
{\cal L}^{\rm eff}_{SI}
&=&\sum_{q=u,d,s}{\cal L}^{\rm{eff}}_q +{\cal L}^{\rm{eff}}_g \ ,
\end{eqnarray}
where
\begin{eqnarray}
{\cal L}^{\rm{eff}}_q
&=& 
f_q \bar{\chi}\chi  ~O_{q}^{(0)} 
+ \frac{g^{(1)}_q}{m_{\chi}} \ \bar{\chi} i  \left(\partial^{\mu}\gamma^{\nu} + \partial^{\nu}\gamma^{\mu}\right)
\chi \  O_{q,\mu\nu}^{(2)}
+ \frac{g^{(2)}_q}{m_{\chi}^2}\
\bar{\chi}(i \partial^{\mu})(i \partial^{\nu})
\chi \  O_{q,\mu\nu}^{(2)}\
,
\label{eff_lagq}
\\
{\cal L}^{\rm eff}_{ g}&=&
f_G \bar{\chi}\chi ~O_{g}^{(0)}
+\frac{g^{(1)}_G}{m_{\chi}}\
\bar{\chi} i \left(\partial^{\mu}\gamma^{\nu} + \partial^{\nu}\gamma^{\mu}\right)
\chi \  O_{g,\mu\nu}^{(2)}
+
\frac{g^{(2)}_G}{m_{\chi}^2}\
\bar{\chi}(i\partial^{\mu}) (i\partial^{\nu})\chi
\
O_{g,\mu\nu}^{(2)} \ .
\label{eq:lag-eft}
\end{eqnarray}
and the SI operators
\begin{eqnarray}
O_q^{(0)} \equiv m_q {\bar q} q \, &,& \quad 
O^{(2)\mu\nu}_{q} \equiv \frac12 \bar{q}\left( \gamma^{\{\mu} iD_-^{\nu\}}  - {g^{\mu\nu} \over 4} i\Dslash_- \right) q \nonumber\\
O_g^{(0)} \equiv G^A_{\mu \nu} G^{A \mu \nu}   \, &,& \quad 
O^{(2)\mu\nu}_g \equiv -G^{A \mu\lambda} G^{A \nu}_{\phantom{A \nu} \lambda} + { g^{\mu\nu} \over 4} (G^A_{\alpha\beta})^2\ .
\end{eqnarray}
The standard shorthand notation used in the above expressions read as,
\begin{eqnarray}
A^{\{\mu}B^{\nu\}} & = &(A^\mu B^\nu + A^\nu B^\mu)/2, \nonumber\\
D^\mu_{\pm} &=&D^\mu \pm \overleftarrow{D}^\mu,  \nonumber \\
D_\mu &=& \partial_\mu - i g A_\mu^A T^A -i e Q A_\mu^A,  \nonumber \\
\overleftarrow{D}_\mu &=& \overleftarrow{\partial}_\mu + i g A_\mu^A T^A +i e Q A_\mu^A .
\end{eqnarray}

 The quantities $f_q,\  g_q^{(1)}$ and $g_{q}^{(2)}$ are Wilson coefficients generated by matrix elements with quarks in the initial and final states, 
 whereas $f_G,\  g_G^{(1)}$ and $g_{G}^{(2)}$ are Wilson coefficients generated by matrix elements with gluons in the initial and final states. 
 Although the operators listed above do not form a  complete basis, they are the set of operators that are relevant and sizable for SI nuclear matrix elements.
 
In this language, the matrix element for dark matter participating in SI
scattering elastically with a target nucleon $(N=\{p, n\})$ is~\cite{Jungman:1995df},
\begin{eqnarray}
f_N/m_N&=&\sum_{q=u,d,s}
 f_{Tq} f_q
+\sum_{q=u,d,s,c,b}
\frac{3}{4} \left[q(2)+\bar{q}(2)\right]\left(g_q^{(1)}+g_q^{(2)}\right)
\nonumber\\
&-&\frac{8\pi}{9\alpha_s}f_{T_G} f_G 
+\frac{3}{4} G(2)\left(g^{(1)}_G
+g^{(2)}_G\right) \ ,
\label{eq:mnuc}
\end{eqnarray}
where $m_N$ is the mass of the nucleon and $f_{Tq}$, $f_{TG}$, $q(2)$, $\bar{q}(2)$ and  $G(2)$ represent hadronic matrix elements:
\begin{eqnarray}
\langle N \vert m_q \bar{q} q \vert N\rangle/m_N  &\equiv& f_{Tq}\ ,  
\nonumber
\\
\langle N \vert -\frac{9 \alpha_s}{8 \pi}  G^A_{\mu \nu} G^{A \mu \nu} \vert N\rangle/m_N  &\equiv& f_{T_G} \ ,  
\nonumber\\
\langle N(p)\vert 
{\cal O}_{q,\mu\nu}^{(2)}
\vert N(p) \rangle 
&=&\frac{1}{m_N}
(p_{\mu}p_{\nu}-\frac{1}{4}m^2_N g_{\mu\nu})\
\left[q(2)+\bar{q}(2)\right] \ ,
\nonumber\\
\langle N(p) \vert 
{\cal O}_{g,\mu\nu}^{(2)}
\vert N(p) \rangle
& =& \frac{1}{m_N}
(p_{\mu}p_{\nu}-\frac{1}{4}m^2_N g_{\mu\nu})\ 
G(2) \ .
\label{eq:nuc-mat}
\end{eqnarray}

The matrix elements of the light quarks $(q= u,\ d,\ s)$ are determined from lattice calculations of the pion nucleon sigma term,
\begin{eqnarray}
\Sigma_{\pi N} &=& \frac{m_u + m_d}{2} \langle N \vert (\bar{u}u + \bar{d}d)\vert N \rangle  \ , \nonumber \\
\Sigma_{-} &=& (m_{d} - m_{u}) \langle N \vert (\bar{u}u - \bar{d}d)\vert N \rangle \ .
\end{eqnarray}	
And the matrix elements of the twist-2 operators are related to the second moments of the parton distribution functions (PDFs):
 \begin{eqnarray}
\left[ q(2)+ \bar{q}(2) \right]&=&\int^{1}_{0} dx ~x~ [q(x)+\bar{q}(x)] \ ,
 \cr
 G(2) &=&\int^{1}_{0} dx ~x ~g(x) \ ,
 \end{eqnarray}
where $q(x)$, $\bar{q}(x)$ and $g (x)$ are the PDFs of quarks, anti-quarks and gluons in $N$, respectively. 
We provide numerical values for the hadronic matrix elements in Appendix~\ref{app:parameters}.

\subsection{Wilson Coefficients}
\label{sec:WilsonCoefficients}

The Wilson coefficients are determined by matching to matrix elements computed in the simplified model.
In this section, we perform this matching at scales of order the mediator mass.  

The leading contributions to the quark Wilson coefficients $f_q,\  g_q^{(1)}$ and $g_{q}^{(2)}$ arise from the tree level diagrams of Figure~\ref{fig:feynqLO},
but at a higher order in expansion of the propagator.
For a single flavor of quark with mass $m_q$ and its corresponding mediator $\tilde{q}$ of mass $M_{\tilde q }$ (and denoted as $M$ in shorthand), the Wilson coefficients are
\begin{eqnarray}
f_q&=&\frac{g_{DM}^2~m_{\chi}}{16(M^2-m_{\chi}^2)^2}
\ ,\nonumber\\
g_q^{(1)}&=&\frac{g_{DM}^2~m_{\chi}}{8(M^2-m_{\chi}^2)^2} \ ,\nonumber\\
g_q^{(2)}&=&0 \ . 
\label{eq:treesq}
\end{eqnarray}
Compared to the SD matrix elements in Equation~\ref{eq:MLO}, 
these Wilson coefficients are suppressed by an additional power of $1/(M^2-m_{\chi}^2)$. Details of the calculation can be found in Appendix~\ref{app:WilsonCoefficients}. We have ommitted the quark mass from the denominators in the expressions above, but use it in our numerical calculations.
\begin{figure}
	\centering
\includegraphics[width=0.65\textwidth]{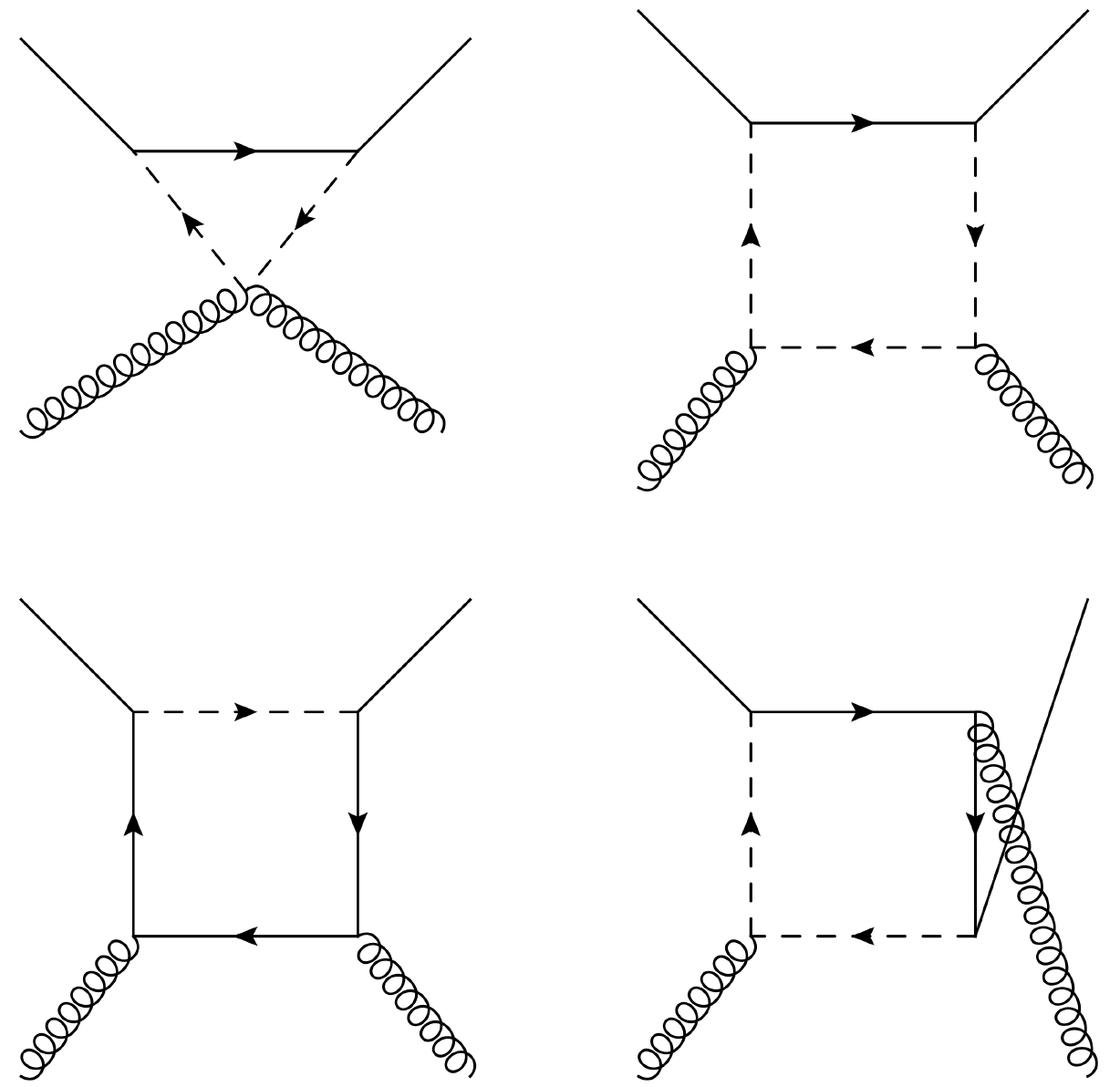}
\caption{Representative Feynman diagrams for DM scattering with gluons. \label{fig:Feynman-gNLO} }
\end{figure}

The leading contribution to the gluonic Wilson coefficients arise at one loop, with representative Feynman diagrams shown in Figure~\ref{fig:Feynman-gNLO} . 
The individual Wilson coefficients are extracted using projection operators, with detailed results relegated to Appendix~\ref{sec:gulonic-full}.
In the limit of small quark mass $(m_q \to 0)$,
\begin{eqnarray}
f_{G} &\simeq & \frac{\alpha_s g_{DM}^2 m_\chi}{192 \pi}\frac{(m_\chi^2 - 2 M^2)}{M^2(M^2 -m_\chi^2)^2},
\label{eq:fGm0} \\
\frac{g_{G}^{(2)}}{m_\chi^2} & \simeq &\alpha_s g_{DM}^2\frac{-2 M^2  m_\chi^2
        +2 \left(M^2- m_\chi^2\right)^2 \log \left(\frac{M^2}{M^2- m_\chi^2}\right)+3
	m_\chi^4}{48 \pi   m_\chi^5 \left(M^2- m_\chi^2\right)^2}
	\label{eq:gG2m0}
\end{eqnarray}
\begin{eqnarray}
\frac{g_{G}^{(1)}}{m_\chi}&\simeq&
\frac{\alpha_s g_{DM}^2}{96 \pi  m_{\chi}^4
	\left(M^2-m_{\chi}^2\right)^2} 
	\bigg[-2 m_{\chi}^4 \log \left( \frac{m_q^2}{M^2} \right)-m_{\chi}^2 \left(M^2+3
m_{\chi}^2\right) \nonumber \\
&+&\left(M^2-3 m_{\chi}^2\right) \left(M^2+m_{\chi}^2\right)
\log \left(\frac{M^2}{M^2-m_{\chi}^2}\right) 
\bigg]
\label{eq:gG1m0}
\end{eqnarray}
They arise at the same power of $1/(M^2-m_{\chi}^2)$, but are suppressed by $\alpha_s$
as compared to the corresponding quark SI Wilson coefficients, cf Eq. \ref{eq:treesq}.  

\begin{figure}
	\centering
	\includegraphics[width=0.65 \textwidth]{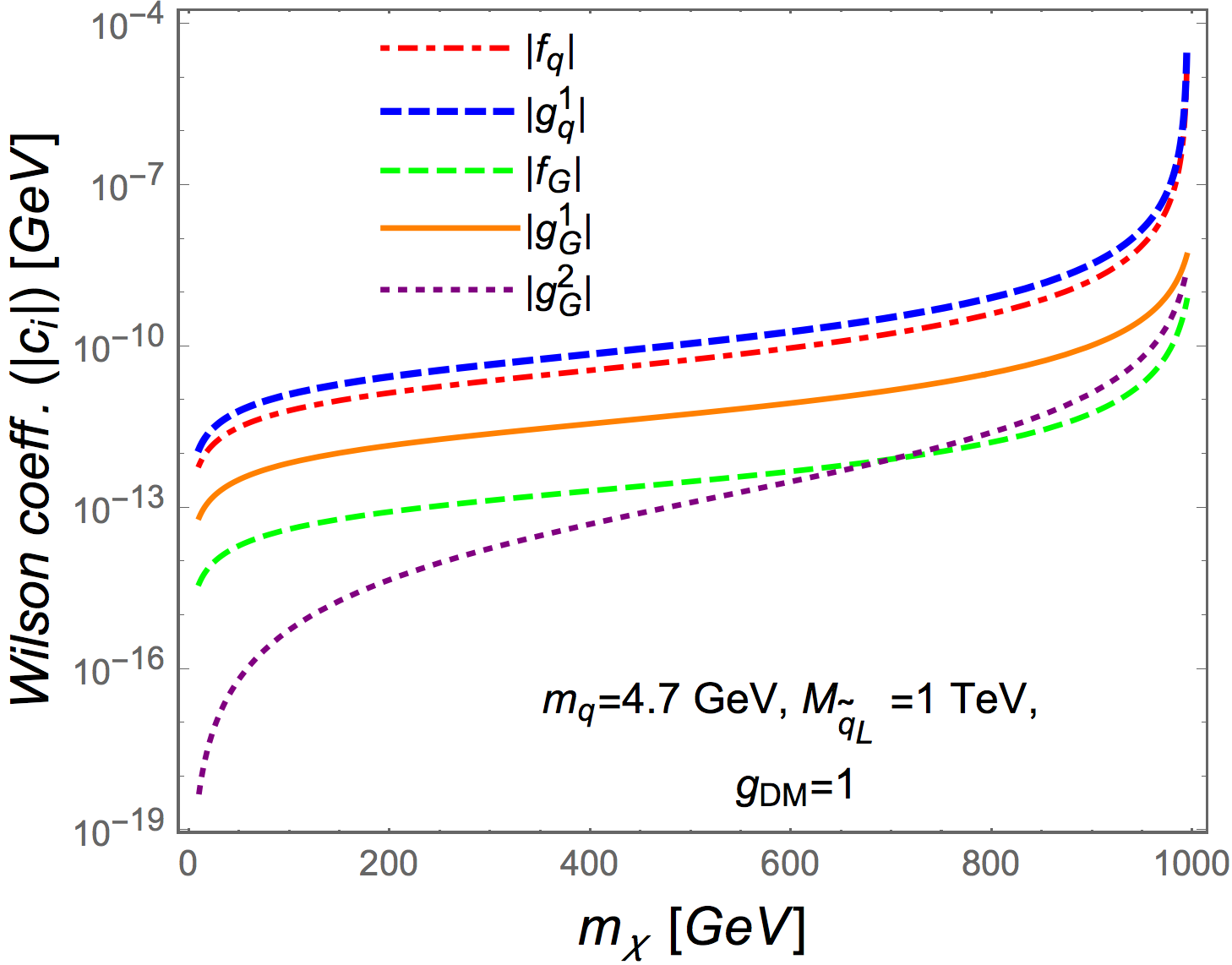}
	\caption{Absolute values of the bottom quark and gluon Wilson coefficients (as indicated, in appropriate powers of GeV for each) 
	as a function of the dark matter mass, for $M_{\tilde{q}} = 1~\text{TeV}$, $m_q= m_b = 4.2 ~{\rm GeV}$ and $g_{DM}=1$.
	\label{fig:plci} }
\end{figure}

In Figure~\ref{fig:plci} we show the absolute value of the bottom quark $({\rm with~ } m_q=m_b= 4.2~\text{GeV})$
and gluon SI Wilson coefficients as a function of the dark matter mass,
and for a representative parameter point with $M_{\tilde{q}} =1~\text{TeV}$ and $g_{DM}=1$.
Each coefficient is expressed in units of GeV to the appropriate power.
All of the Wilson coefficients have a  resonant enhancement in the limit $m_{\chi} \to M_{\tilde{q}}$. 
We observe that the gluonic Wilson coefficients are roughly an order of magnitude smaller than their quark counterparts. 
However, this feature is mitigated by the fact that their hadronic matrix elements are large and RGE effects are important, 
especially for the spin-0 gluonic term $\mathcal{O}_{g}^{(0)}$. 

We note that while the quark coefficients $f_q$ and $g_{q}^{(1)}$ are independent of the quark mass at this order, the gluonic coefficients
$f_G$, $g_G^{(1)}$ and $g_{G}^{(2)}$ all depend on the mass of the quark in the loop.  In the limit of $m_q \to 0$,
$f_G$ and $g_{G}^{(2)}$ reduce to the finite expressions in Equations~(\ref{eq:fGm0}) and (\ref{eq:gG2m0}).  
In contrast, as indicated in Equation~(\ref{eq:gG1m0}), $g_G^{(1)}$ diverges logarithmically with divergent piece:
\begin{equation}
\Delta g_G^{(1)} = \frac{\alpha_s g_{DM}^2m_\chi}{24 \pi (M^2 - m_\chi^2)^2}\log\left(\frac{M}{m_q}\right)
= g_{q}^{(1)}~\frac{\alpha_s }{3 \pi }~\log\left(\frac{M}{m_q}\right) \ .
\label{eq:collineardiv}
\end{equation}
From the second expression, we observe that it can be rewritten in terms of the quark Wilson coefficient $g_q^{(1)}$,
suggesting that it might cancel against the $\alpha_s$ real correction to the quark scattering.
We demonstrate below in the context of the renormalization group evolution that this is indeed the case.

\subsection{Renormalization Group Evolution and Threshold Matching}
\label{sec:wil}

The Wilson coefficients are matched to the simplified model 
at a scale $\mu \simeq M_{\tilde{q}}$ which is appropriate to describe production of the mediators at the LHC.  To
make accurate predictions at the low energy scales appropriate for direct detection, we evolve them to $\mu_l = 2$~GeV via 
renormalization group equations, which we evaluate at leading log in the strong coupling $\alpha_s$, following
Refs.~~\cite{DEramo:2014nmf,Hill:2014yka,Hill:2014yxa}.  We neglect subleading corrections from the electroweak
interactions.

The strong force corrections to the SI 
operators boil down to the corrections to the quark and gluon
bilinear factors $O_q^{(0)}$, $O^{(2)\mu\nu}_{q}$, $O_g^{(0)}$, and $O^{(2)\mu\nu}_g$ in the SI 
EFT Lagrangian\footnote{The SD EFT operators contain $V_q^\mu= \bar{q} \gamma^\mu q $ and $A_q^\mu=\bar{q} \gamma^\mu \gamma_5 q$,
	which do not receive large RGE corrections.} of Equation~\ref{eq:lag-eft}, Equation~(\ref{eq:lag-eft}).  Under the renormalization group,
the operators $O_i$ and their Wilson coefficients $c_i$ evolve according to their anomalous dimensions $\gamma_{ij}$:
\begin{align}
{d\over d\log\mu} O_i = - \gamma_{ij} O_j \,, \quad
{d\over d\log\mu} c_i = \gamma_{ji} c_j \,.
\label{eqn:evol}
\end{align}
The solution evolving from a high scale $\mu_h$ down to a low scale $\mu_l$ takes the form:
\begin{equation}
c_{i}(\mu_l) = R_{ij}(\mu_l,\mu_h)c_{j}(\mu_h)\ ,
\end{equation}
where $R$ is a square matrix in flavor space and $c_i$ and $c_j$ are  column vectors of Wilson coefficients arranged in flavor space as  $c_{j} = \left(u,d,s,c,b,t|g\right) $ . Conservation of angular momentum forbids
mixing between the scalar $(0)$ and tensor $(2)$ operators, allowing us to consider them
in two separate groups.  For each group, the matrix $R^{(i=0,2)}$ is square in flavor space with $n_f$ quark flavors:
\renewcommand{\arraystretch}{1.4}
\begin{align}\label{eq:R1}
R^{(i)}
&=  \left( \begin{array}{ccc|c} 
& & & R^{(i)}_{qg} \\
&  \mathbb{I} (R^{(i)}_{qq} - R^{(i)}_{qq^\prime} ) + \mathbb{J}  R^{(i)}_{qq^\prime} & & \vdots \\
&&  & R^{(i)}_{qg} \\
\hline
R^{(i)}_{gq} & \cdots & R^{(i)}_{gq} & R^{(i)}_{gg}
\end{array} 
\right) \,,
\end{align}
where the $n_f \times n_f$ matrices $\mathbb{I}$ and $\mathbb{J}$ are the identity matrix and the matrix with all elements equal to unity, respectively.
The upper $n_f \times n_f$ block diagonal entries describe mixing among the quark flavors, and lower diagonal entry renormalize the gluonic
operator.  The block-off-diagonal terms induce mixing between the quark and gluon operators.
For the scalar operators~\cite{Hill:2014yxa}:
\begin{align}
& R^{(0)}_{qq} =1,  & R^{(0)}_{qg} &= 2 [\gamma_m (\mu_h) - \gamma_m(\mu_l)]/{\tilde \beta}(\mu_h) \, ,\ \nonumber\\
& R^{(0)}_{qq^\prime} = R^{(0)}_{gq} = 0 \, , & R^{(0)}_{gg} &= {\tilde \beta}(\mu_l)/ {\tilde \beta}(\mu_h)  \ ,  
\end{align}
where $\beta$ is the QCD beta function and $\tilde{\beta} \equiv \beta/g_s$.
At this order, there is no mixing between operators of different quark flavors. 
The form of the quark/gluon mixing can be understood as a sum rule that determines the quark and gluon contributions to the mass of the nucleon. 
For the spin-2 operators~\cite{Hill:2014yxa}:
\begin{align}
& R^{(2)}_{qq} = R^{(2)}_{qq^\prime} + r(0) \,, & R^{(2)}_{qq^\prime} &=  {1 \over n_f} \Big[ {16 r(n_f) +3 n_f \over 16 + 3n_f} - r(0) \Big] \, , \nonumber\\
& R^{(2)}_{qg} = {16[1-r(n_f)] \over 16 + 3n_f}  \, , & R^{(2)}_{gq}  &= {3[1-r(n_f)] \over 16 + 3n_f} \, , \nonumber\\
& R^{(2)}_{gg} =  {16+3n_f r(n_f) \over 16 + 3n_f} \ ,
\end{align}
where
\begin{equation}
r(t) \equiv \left(\alpha_s(\mu_l) \over \alpha_s(\mu_h) \right)^{ -{1\over 2\beta_0} \left( {64\over 9}  + \frac43 t \right) }\, ,
\end{equation}
with $\beta_0 = 11 - 2/3 n_f$.  For the spin-2 operators, there is mixing both between quark flavors and between quarks and the gluons. They are 
governed by sum rules of PDFs in nucleons.

At the threshold scale for each heavy quark $\mu= m_{Q}$, the heavy quark is integrated out pertrubatively and the Wilson coefficients in the $n_f+1$ flavor theory
are matched to the $n_f$ flavor theory:
\begin{equation}
c_i^{(n_f)}(\mu_Q) = {M}_{ij}(\mu_Q) c^{(n_f +1)}_j (\mu_Q) \,,
\end{equation}
where $M_{ij}$ is the rectangular matrix
\renewcommand{\arraystretch}{1.2}
\begin{align}\label{eq:Mlead}
M^{(i)} = \left( \begin{array}{ccc|c|c} 
1 &  & & 0 & 0 \\
&  \ddots  & &\vdots & \vdots \\
&  &  1~ & 0 & 0 \\
\hline
0 & \cdots & 0~ & ~M^{(i)}_{gQ}~ & ~M^{(i)}_{gg} 
\end{array} 
\right) \, ,
\end{align}
and $n_f $ denotes the number of light quark flavors with quark mass less than the energy scale $\mu_Q$.
The entries of the matching matrix for the spin-0 operators are~\cite{Hill:2014yxa}:
\begin{align}
&M^{(0)}_{gQ} = -{\alpha_s^{\prime}(\mu_Q)\over 12\pi} 
\Big\{ 1 + {\alpha_s^{\prime}(\mu_Q) \over 4\pi} \left[  11 - \frac43 \log {\mu_Q \over m_Q} \right]  +\order(\alpha_s^2) \Big\} \, ,& \nonumber \\
&  M^{(0)}_{gg} =1 - {\alpha_s^{\prime}(\mu_Q) \over 3\pi} \log {\mu_Q \over m_Q}   +\order(\alpha_s^2) \ .&
\end{align}
Here $\alpha_s^{\prime}$ denotes the strong coupling in the $n_f + 1$ flavor theory.
The elements of the matching matrix for spin-2 operators are~\cite{Hill:2014yxa}:
\begin{align}
& M^{(2)}_{gQ} =  {\alpha_s^{\prime} \over 3\pi}  \log {\mu_Q \over m_Q}  + \order(\alpha_s^2) , & \nonumber \\  
& M^{(2)}_{gg} = 1 + \order(\alpha_s) . &
\end{align}
By matching at the scale $m_Q$, only $M_{gQ}^{(0)}$ results in a non-trivial correction.

\begin{figure}
	\centering
	\includegraphics[width=0.65 \textwidth]{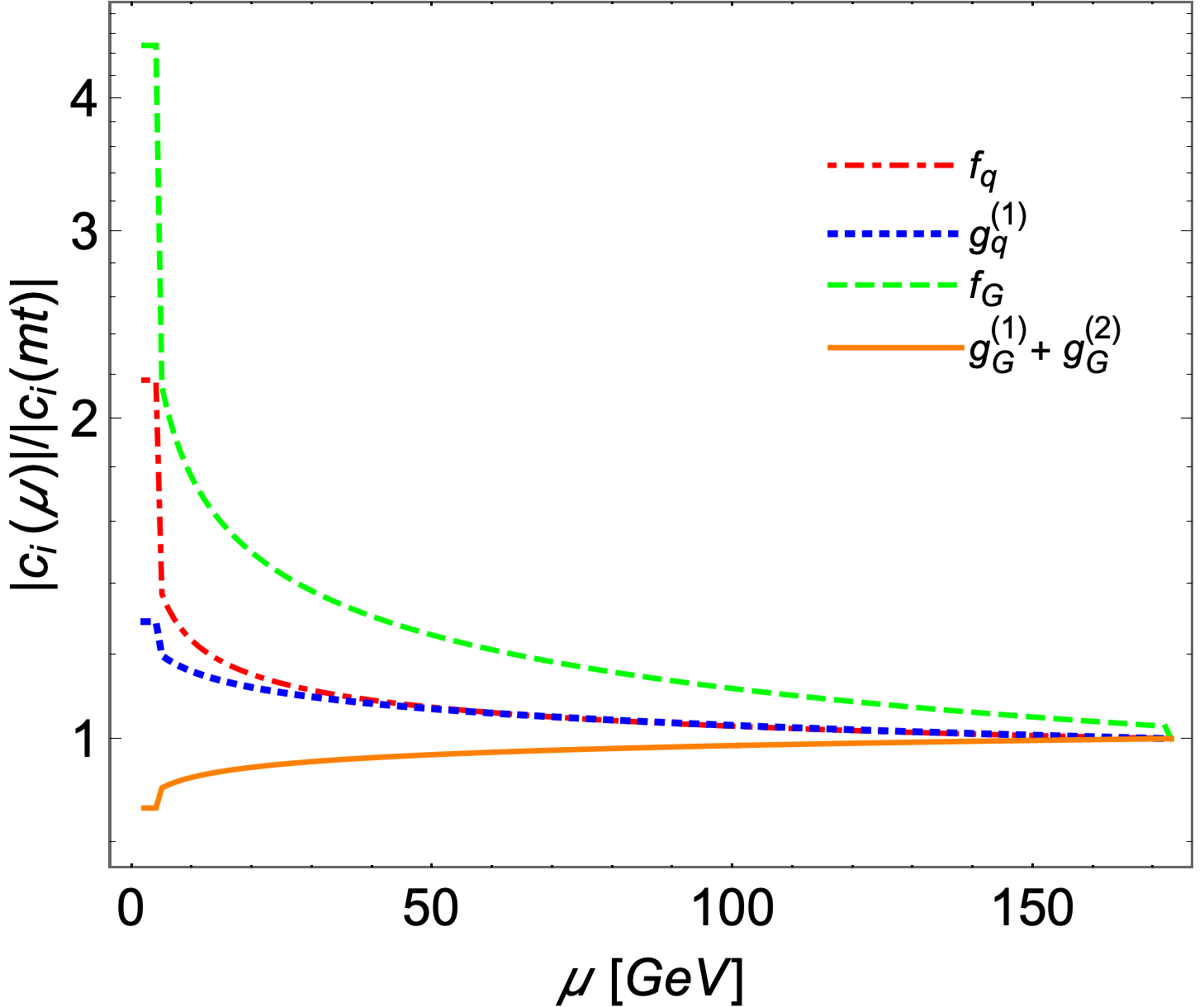}
	\caption{Variation of the $\log_{10}$ ratio of Wilson coefficients $c_{i}(\mu)/c_{i}(m_t)$, as indicated, with the scale $\mu$, for the $Q_L$ model with $m_\chi = 10$~GeV, 
	$M_{\tilde{q}}=1$~TeV, and $g_{DM}=1$. \label{fig:rge1}
	}
\end{figure}
In Figure~\ref{fig:rge1}, we show the ratio of the Wilson coefficients to their values at the top quark mass, $c^{(n_f)}_{i}(\mu)/c^{(6)}_{i}(m_t)$, as a function of scale for the 
$Q_L$ model with $m_\chi = 10$~GeV, $M_{\tilde{q}}=1$~TeV and $g_{DM}=1$.  At the lowest scale we consider, $\mu=2~{\rm GeV}$, the largest impact is on
the spin-0 operators, whose Wilson coefficients change by a factor of $\sim 2 - 5$, with a more modest impact on the spin-2 operators  -- 
$g_{q}^{(1)}$ increases by a factor of $\sim 1.1$ whereas $(g_{G}^{(1)} +g_{G}^{(2)})$  reduces by a small  factor of $\sim 0.87$. We have left out $g_{q}^{(2)}$, since it is zero,
and we group  the gluonic twist-2 Wilson coefficients together, since they appear as sums in the nucleon matrix element in Equation~(\ref{eq:mnuc}).
Although both $f_{G}$ and $f_{q}$ increase at low scales, since the hadronic matrix element for $f_q$ is much smaller compared to the other terms, it does not
have much impact on the total nucleon dark matter scattering amplitude. 
The net effect on the amplitude is that, at low scales, it increases roughly by a factor of $\sim 2$ when compared to high scales.

\subsubsection{Collinear Divergence}
\label{sec:thelogs}

We observed above that the one loop Wilson coefficient $g_{G}^{(1)}$ contains a collinear divergence, Equation~(\ref{eq:collineardiv}).
This divergence is canceled to order $\alpha_s$ by the RGE contribution to $g_G^{(1)}$ from $g_q^{(1)}$.  We illustrate how this works for
the bottom quark contribution in the $d_R$ model when the energy scale reduces from $\mu_h$ to $\mu_l$ with $\mu_h > m_b$~\footnote{This works for all the $q_L$ and $u_R$ models as well, where the first threshold occurs at the top quark mass. For the $d_R$ model the first threshold in the wilson coefficients occurs at $\mu=m_b$ and the usual threshold at $\mu = m_t$ still exists in the strong coupling $\alpha_s$.}.
To expand the RGE contribution, we note that the ratio $\alpha_s (\mu_h) / \alpha_s (\mu_l)$ can be written as:
\begin{equation}
\frac{\alpha_s(\mu_h)}{\alpha_s(\mu_l)} = 1 + \frac{\alpha_s(\mu_h)\beta_0}{2\pi}\log\left[\frac{\mu_l}{\mu_h}\right] ,
\end{equation}
where $\beta_0= 11 -2/3 n_f$, which implies that the factor $r(t)$ in the RGE is
\begin{eqnarray}
	r(t) &=& \left(\alpha_s(\mu_l) \over \alpha_s(\mu_h) \right)^{ -{1\over 2\beta_0} \left( {64\over 9}  + \frac43 t \right) } 
	\simeq 1 + \frac{ \left( {64\over 9}  + \frac43 t \right) \alpha_s(\mu_h)}{4\pi}\log\left[\frac{\mu_l}{\mu_h}\right] .
\end{eqnarray}
Expanding the RGE contribution to $g_G^{(1)}$ from $g_q^{(1)}$ and combining with the collinear divergent term Equation.~(\ref{eq:collineardiv}) yields:
\begin{eqnarray}
\Delta g_G^{(1)}\bigg|_{\mu_l} &\simeq& \frac{m_{\chi}g_{DM}^2}{72 \pi^2(M_{\tilde{q}}^2-m_{\chi}^2)^2}\bigg[3 \pi  \alpha_s(\mu_h) \log
	\left(\frac{\mu_l}{\mu_h}\right)\nonumber \\
	&+&\alpha_s(M_{\tilde q}) \log
	\left(\frac{M_{\tilde q}}{m_b}\right) \left(3 \pi -5
	\alpha_s(\mu_h) \log \left(\frac{\mu_l}{\mu_h}\right)\right)\bigg].
\end{eqnarray}
To order $\alpha_s$, the collinear logs cancel provided one chooses $\mu_h = M_{\tilde q }$ and $\mu_{l}= \mu_b$.  This procedure removes the large
log dependence for the heavy quarks.  For the light quarks ($u, d, s$), whose masses are below the hadronic matching scale $\mu_l = 2$~GeV, 
the cancellation works as outlined above down to $\mu_l = 2~ {\rm GeV}$, with the remaining portion fo the divergence being absorbed into their $\overline{\text{MS}}$ masses at that scale.

\subsection{Limits from Direct Searches}

\subsubsection{Spin Dependent Limits}
\label{sec:SD}

\begin{figure}
	\centering
	\includegraphics[width=0.475 \textwidth]{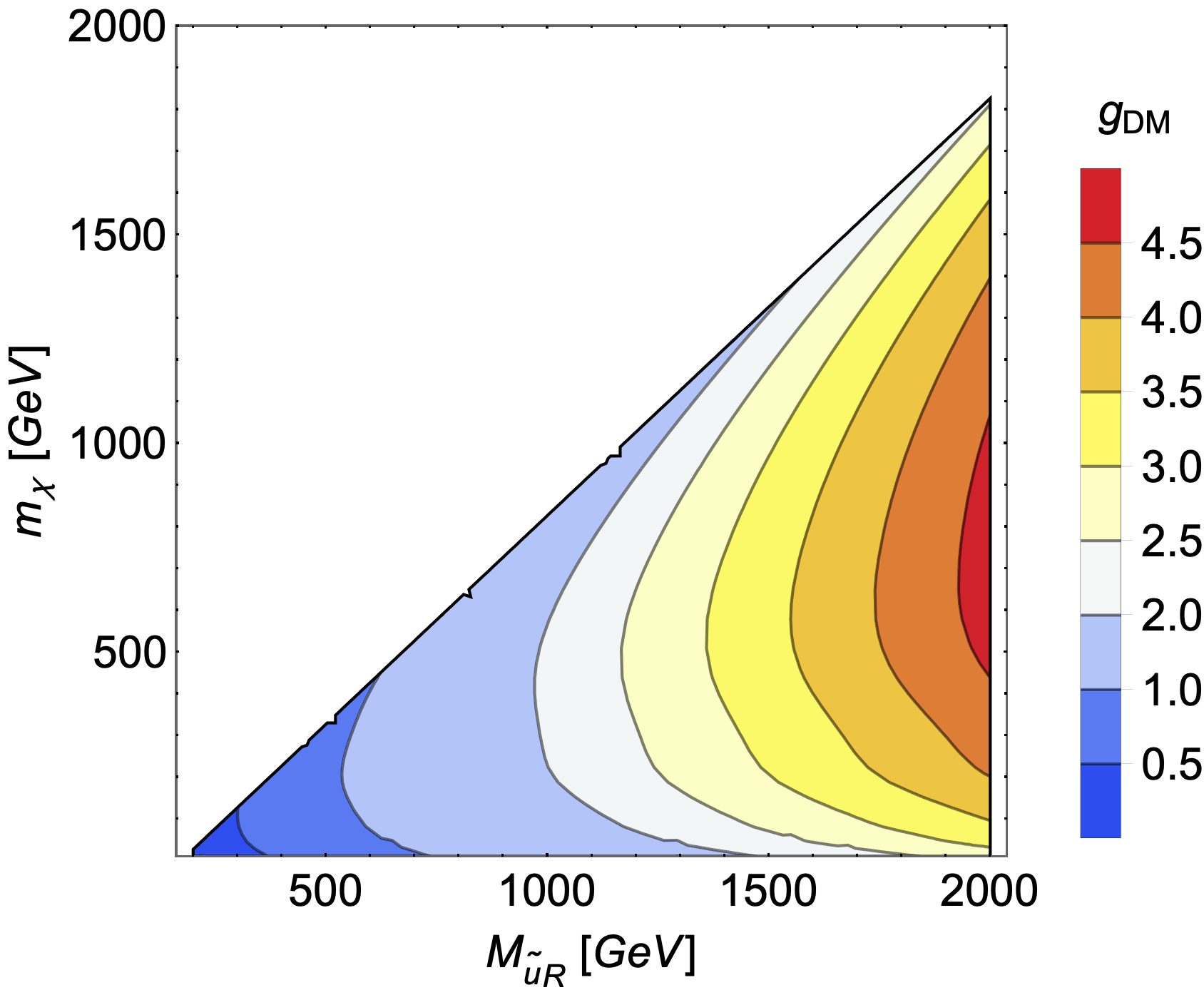}
	\includegraphics[width=0.475 \textwidth]{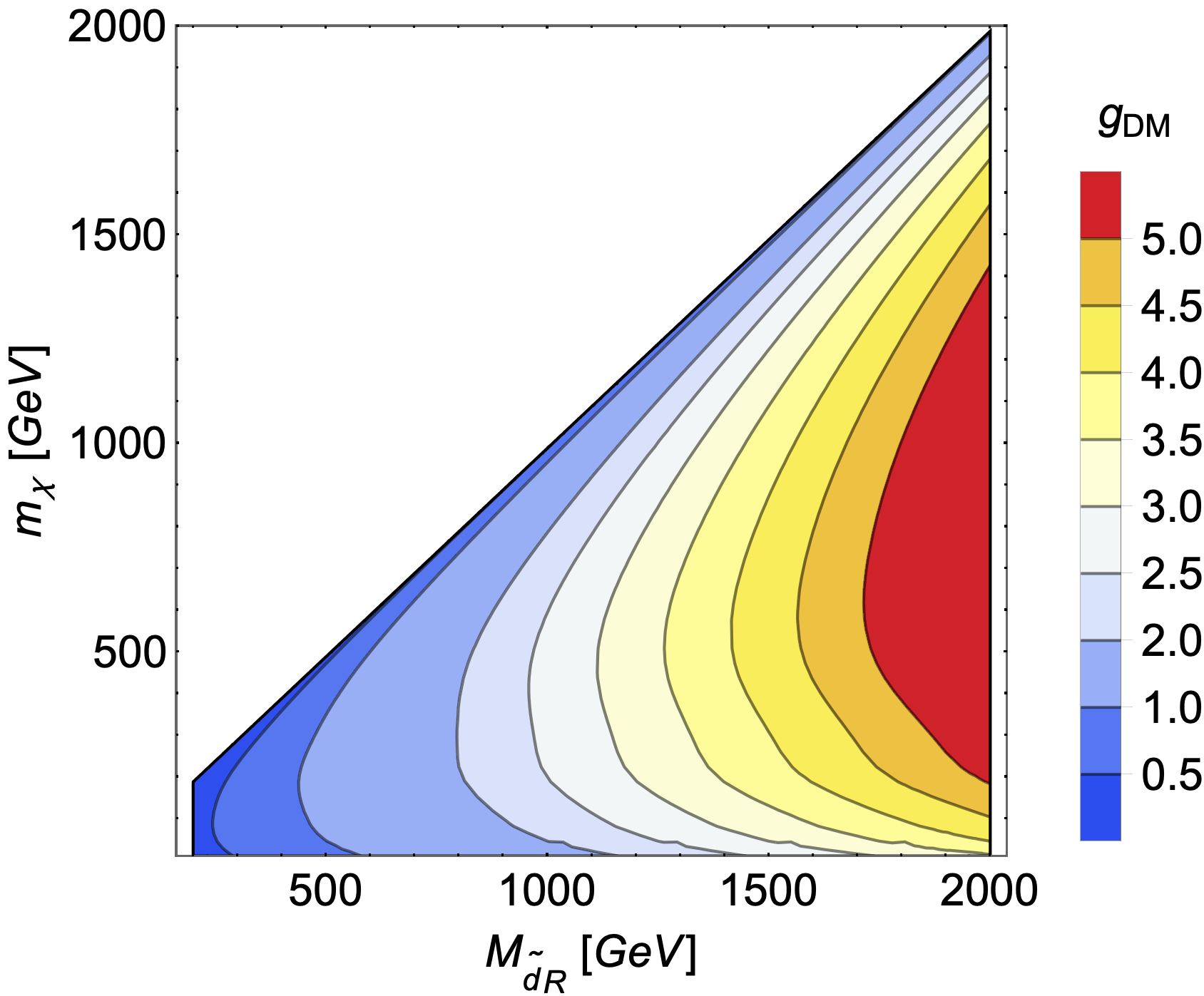}
	\includegraphics[width=0.475 \textwidth]{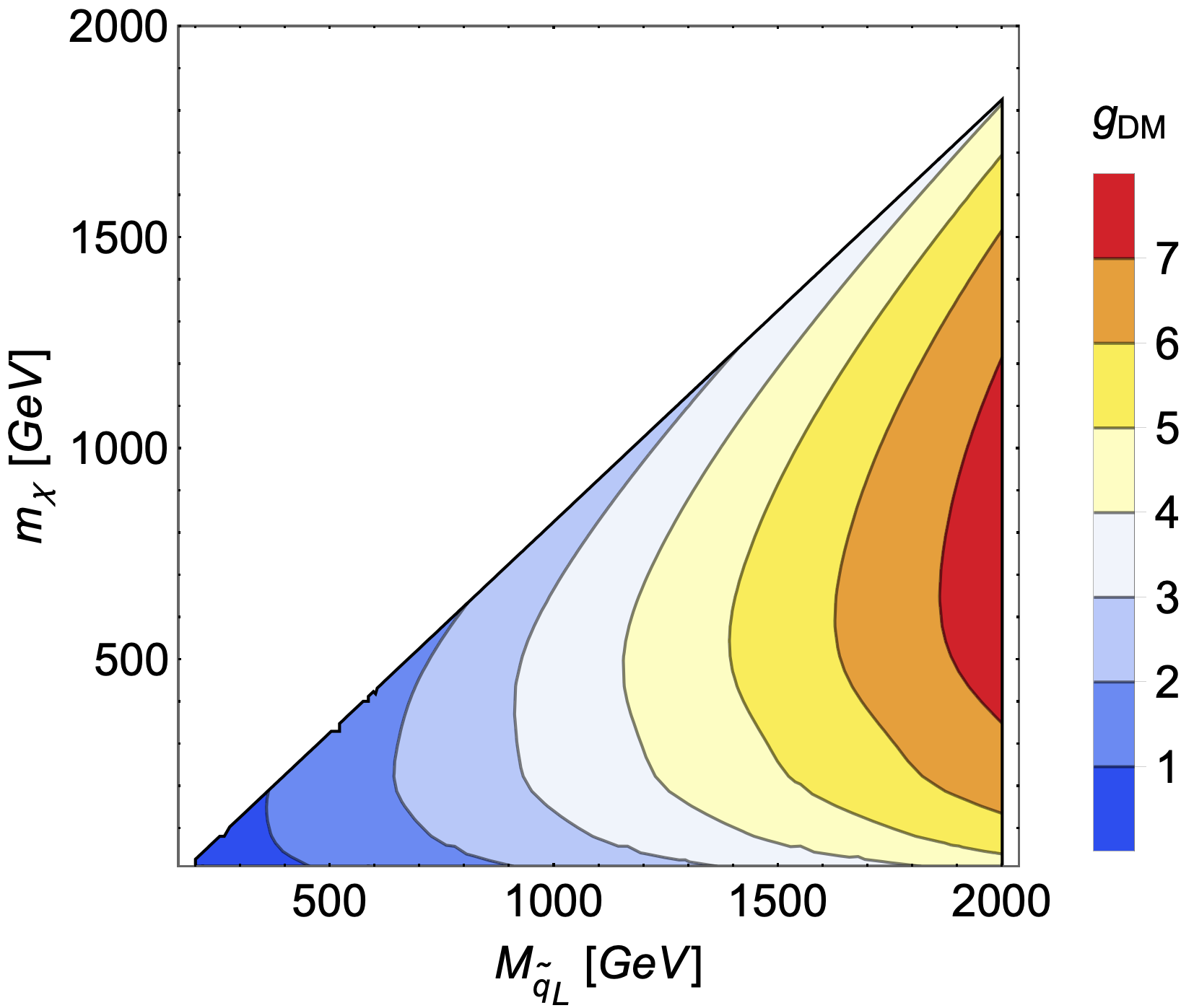}
	\caption{Limits on $g_{DM}$ in the plane of the dark matter and mediator masses from the SD cross section for the
	 $u_R$ (upper left), $d_R$ (upper right) and $q_L$ (bottom) models. 
	 	\label{fig:SDlim1}
	}
\end{figure}

The SD cross section is dominated by its tree level contribution at leading order in the $1/M_{\tilde{q}}$ expansion.  
A detailed discussion of the matching to the hadronic EFT can be found in Ref.~\cite{Freytsis:2010ne}, ands results
in the SD cross sections for the $u_R$, $d_R$, and $Q_L$ models~\cite{DiFranzo:2013vra}:
 \begin{align}
\sigma_{SD}^{u_R} &= \frac{3}{16\pi} \frac{m_N^2M_{\chi}^2}{(m_N+M_{\chi})^2} \frac{g_{\!_{DM}}^4}{(M_{\tilde{d}}^2-M_{\chi}^2)^2} (\Delta u^N)^2, \\
\sigma_{SD}^{d_R} &= \frac{3}{16\pi} \frac{m_N^2M_{\chi}^2}{(m_N+M_{\chi})^2} \frac{g_{\!_{DM}}^4}{(M_{\tilde{d}}^2-M_{\chi}^2)^2} (\Delta d^N+\Delta s^N)^2, \\
\sigma_{SD}^{q_L} &= \frac{3}{16\pi} \frac{m_N^2M_{\chi}^2}{(m_N+M_{\chi})^2} \frac{g_{\!_{DM}}^4}{(M_{\tilde{d}}^2-M_{\chi}^2)^2} (\Delta u^N+\Delta d^N+\Delta s^N)^2,
\label{eq:SDcs}
\end{align}
where $\Delta u^N, \Delta d^N $ and $\Delta s^N$ are matrix elements, whose values are tabulated in Appendix~\ref{app:parameters}.  Because the spin of a
heavy nucleus is typically dominated by a single unpaired nucleon, various direct search 
experiments are typically more sensitive to either scattering with a proton or a neutron, depending on the target nucleus.  Currently, the best SD limits on
SD proton scattering for $m_\chi \gtrsim 3.5$~GeV are from PICO-60~\cite{Amole:2017dex} (and from CDMSlite, below that~\cite{Agnese:2017jvy}), and 
the best limits on SD neutron scattering are from LUX~\cite{Akerib:2017kat}.  
For the all three of the simplified models under consideration, the most stringent constraints are from SD proton scattering.
In Figure~\ref{fig:SDlim1} we show the constraints that arise from the PICO-60 limits
for each of the three simplified models,
in the plane of the dark matter and mediator masses, with the colored shading representing the upper limit on $g_{DM}$ consistent with the null search results.
White regions indicate where the mediator mass is smaller than the dark matter mass.
Over-all, the constraints are very weak, $g_{DM} \lesssim 5$, generically allowing any perturbative value of $g_{DM}$ for mediator masses  greater than a couple of TeV,
although they are somewhat stronger in the resonant region $M_{\tilde{q}} \sim m_\chi$ and for small dark matter mass.

\subsubsection{Spin Independent Limits}
\label{sec:SI}

The cross section for SI scattering with a nucleon is expressed in terms of the form factors $f_N= \{f_p, f_n \}$ as:
\begin{align}
\sigma^{N}_{SI} = 
\frac{4}{\pi}\left(\frac{m_\chi ~ m_N}{M +m_N}\right)^2
\left|f_N\right|^2\ ,
\end{align}
where $f_N$ is related to the Wilson coefficients via Equation~(\ref{eq:mnuc}).  We find that comparing the RGE-improved to the non-RGE-improved
results for a typical point in parameter space, the RGE-improved $f_N$ are generally 
about a factor of $\sim 1.9$ larger than the non-RGE-improved results.  This translates into an increase of about a factor of 
$\sim 4$ in the SI cross section, and highlights the importance of the higher order terms to accurately assess $\sigma_{SI}$ in this simplified model.

\begin{figure}
	\centering
	\includegraphics[width=0.475 \textwidth]{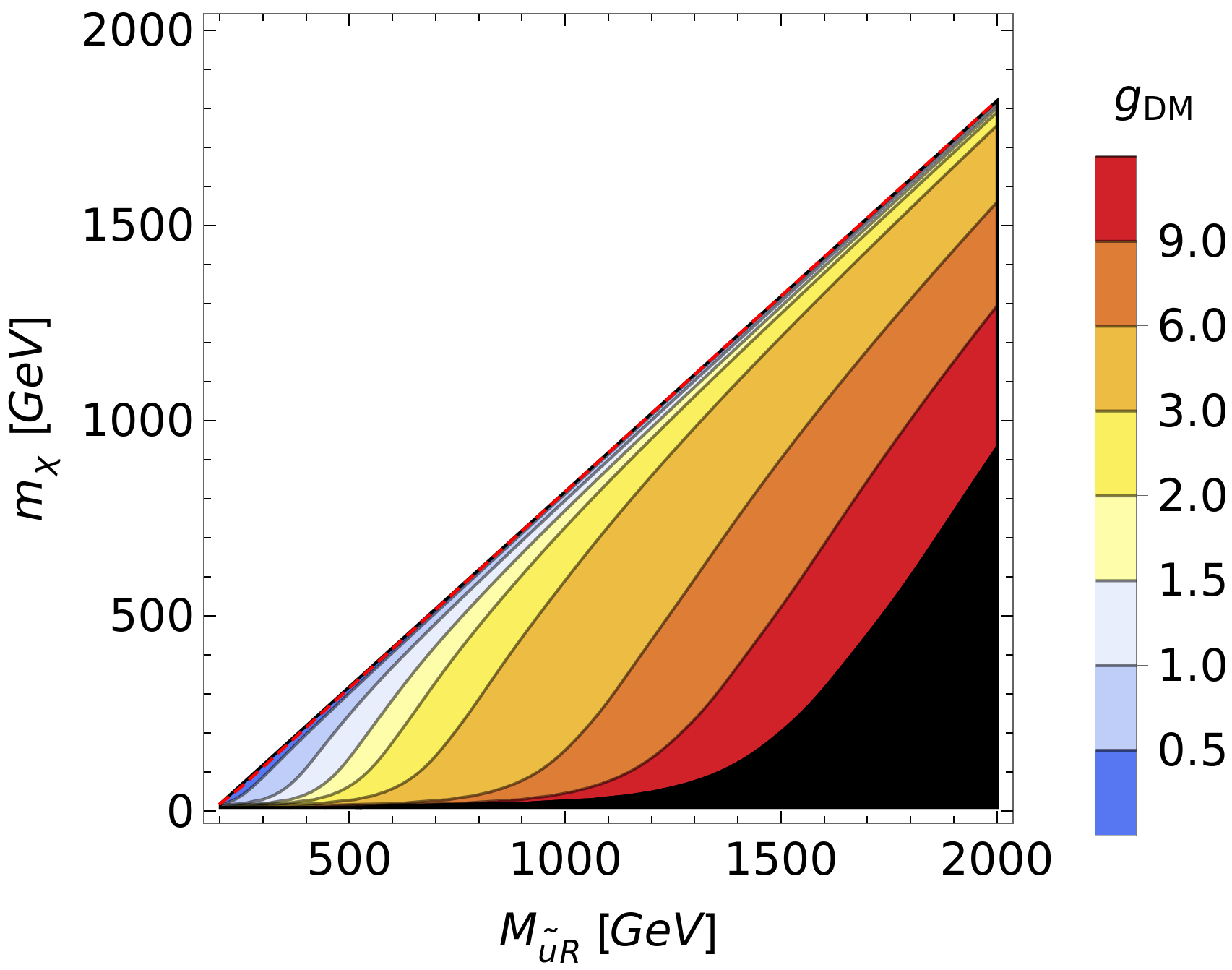}
	\includegraphics[width=0.475 \textwidth]{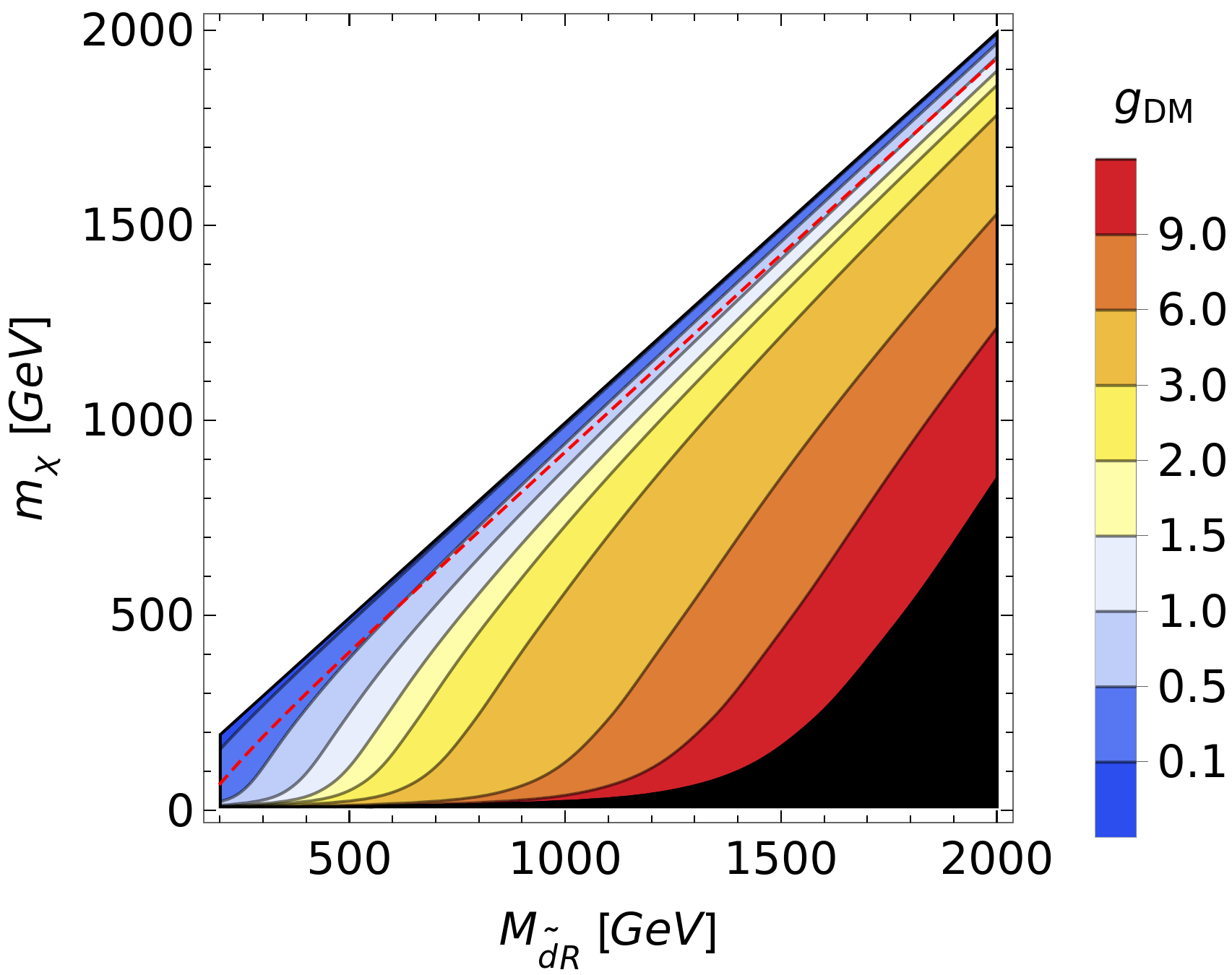}
	\includegraphics[width=0.475 \textwidth]{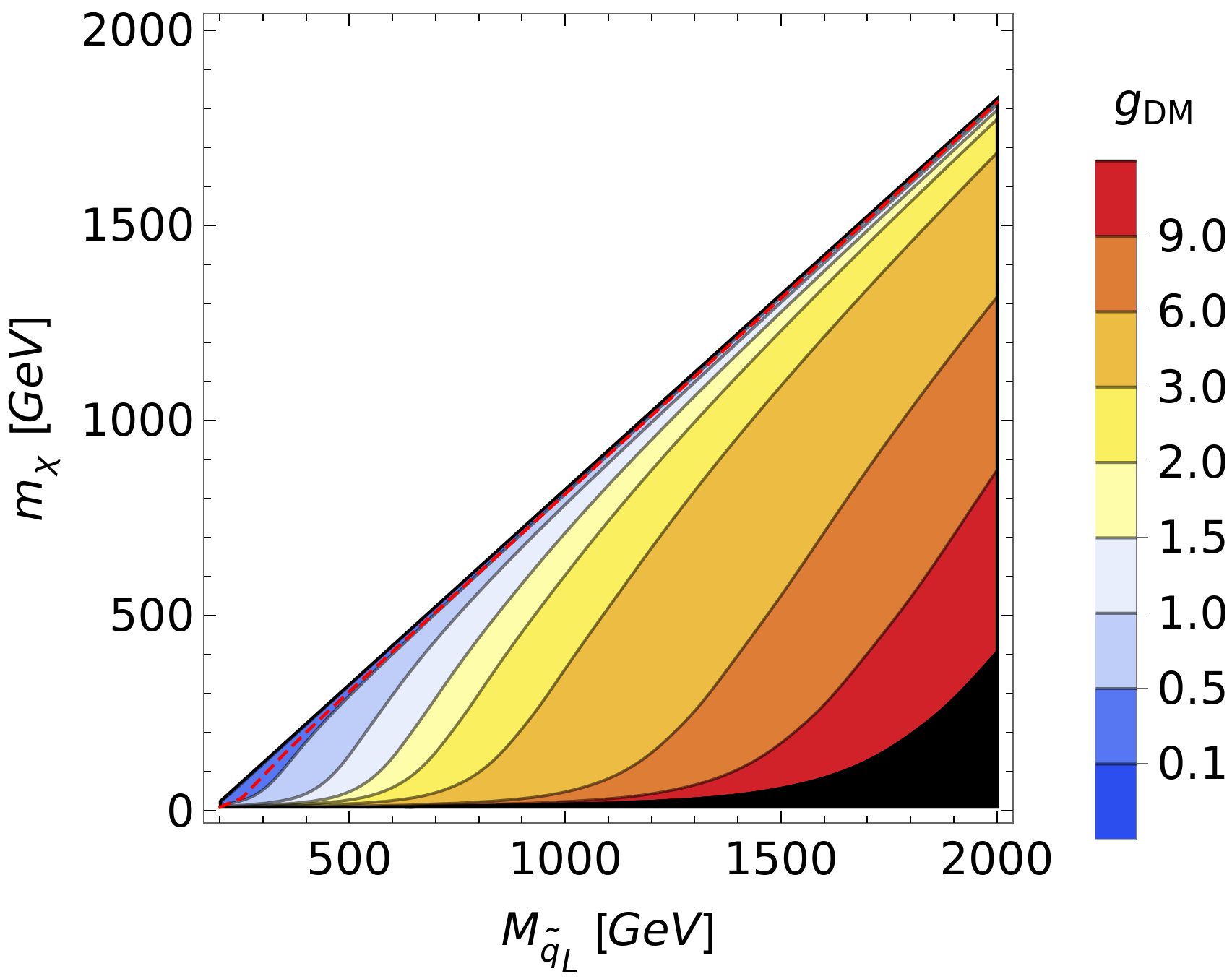}
	\caption{Limits on $g_{DM}$ in the plane of the dark matter and mediator masses from the SI cross section for the
	 $u_R$ (upper left), $d_R$ (upper right) and $q_L$ (bottom) models. 
	 The red dashed line indicates the region of parameter space above which the partial width for the mediators is less than $\Lambda_{QCD}$.
	\label{fig:SI_DD}}
\end{figure}

Experiments sensitive to SI scattering typically unfold the dominant nuclear physics to place limits on $\sigma^{N}_{SI}$, with the best current limits
for masses above a few GeV obtained from the null searches of the Xenon 1T experiment~\cite{Aprile:2017iyp}.
In Figure~\ref{fig:SI_DD} we show limits on $g_{DM}$ for the $(q_L,u_R,d_R)$ models in the plane of the dark matter and mediator masses. 
The shaded regions correspond to allowed values of $g_{DM}$.  Comparing with Figure~\ref{fig:SDlim1}, we observe that despite being 
suppressed because they arise at higher orders, the SI limits are stronger (typically by about an order of magnitude)
than the SD limits especially in the region of parameter spaces when  $m_\chi \simeq M_{\tilde{q}}$ with $g_{DM} \lesssim 0.1$ in the resonant region.

\section{Dark Matter Production at the LHC}
\label{sec:constraints}

\begin{figure}
	\centering
	\includegraphics[scale=0.12]{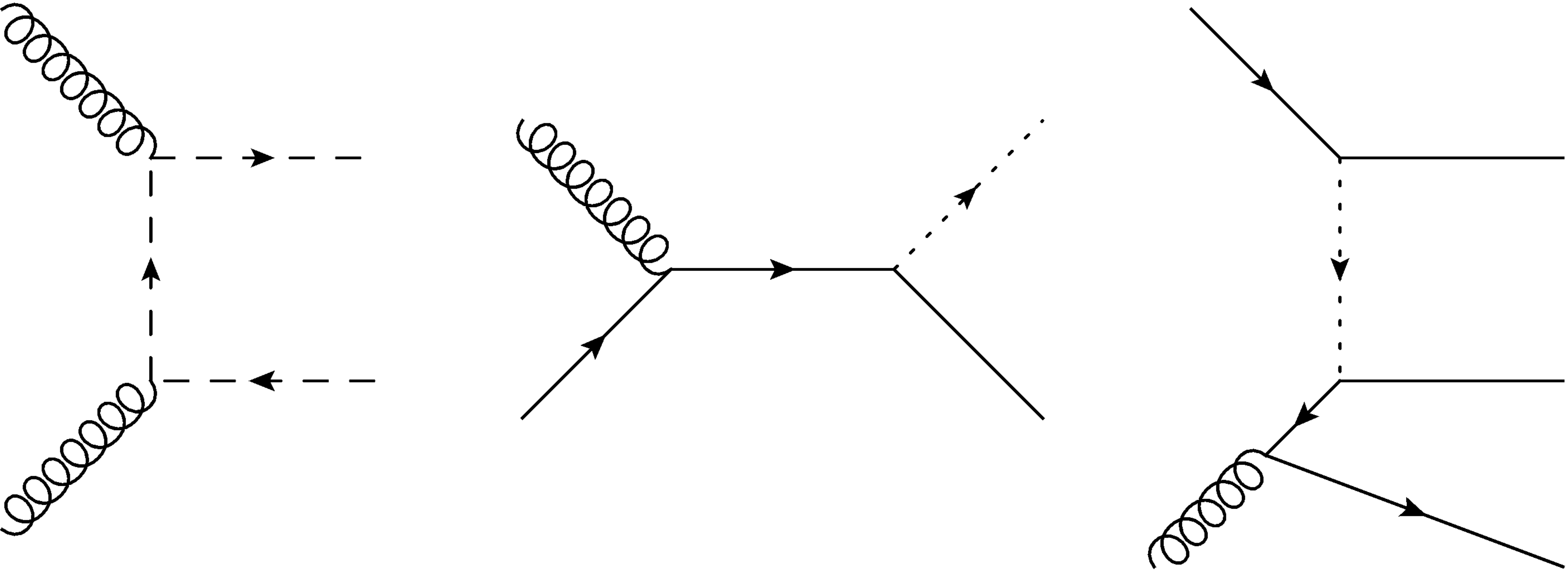}
	\caption{Representative Feynman diagrams for
	(left to right): mediator pair production, associated production, and $\chi \chi$ + jet production at the LHC.}
	\label{sqpair}
\end{figure}

At the LHC, there are three short distance processes of interest, as predicted by this model, which lead to missing transverse momentum ($\slashed{E}_{T}$) signatures:
\begin{itemize}
\item pair production of the colored mediators $(\tilde q)$, followed by their decay into dark matter $(\chi)$ plus a quark;
\item associated production of $\tilde{q}$ with $\chi$; and
\item pair production of the dark matter in association with a jet from initial state radiation, $p p \rightarrow \chi \chi j$.
\end{itemize}
The first processes are dominated by production of the mediators through the strong force (for $g_{DM} \ll g_s$) yielding a jets + $\slashed{E}_{T}$ signature.
A special case has a top-flavored mediator, whose decay into a top quark often also results in charged leptons and bottom-flavored quarks in the final state.
The rate of the associated
production of $\tilde q $ and $\chi$ and dark matter pair production processes are controlled by $g_{DM}$.  For the parameter space of interest, the dark matter pair production process is always
subdominant, and will be neglected from here on.

\subsection{Cross-sections at NLO}
\label{sec:totcs}

Robust interpretations of LHC data require comparison with precise theoretical determinations of cross sections.  We compute the inclusive rates for
the mediator pair and associated production processes at NLO in QCD.
These calculations are performed in the {\tt MG5\_aMC@NLO} framework~\cite{Alwall:2014hca}, with the simplified models implemented via 
{\tt FeynRules}~\cite{Alloul:2013bka}. One-loop corrections are computed in {\tt MadLoop}, based on the OPP method~\cite{Ossola:2006us}, 
with ultraviolet divergences and rational $R_2$ terms
via the {\tt NLOCT} package~\cite{Degrande:2014vpa}.  
CT14NNLO PDF sets are employed in both LO and NLO QCD results~\cite{Dulat:2015mca},
with factorization and renormalization scale chosen to be 
$\mu_f=\mu_r=\dfrac{1}{2}H_T=\dfrac{1}{2}\sum_i\sqrt{p_{T,i}^2+m_i^2}$, where the sum is over all of the final-state particles. 

\begin{figure}\centering
	\includegraphics[width=0.45\linewidth]{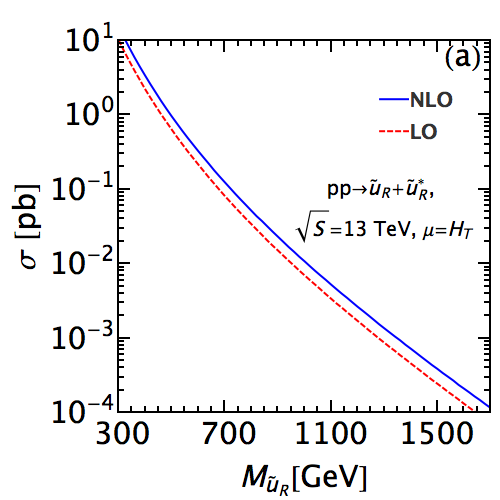}
	\includegraphics[width=0.45\linewidth]{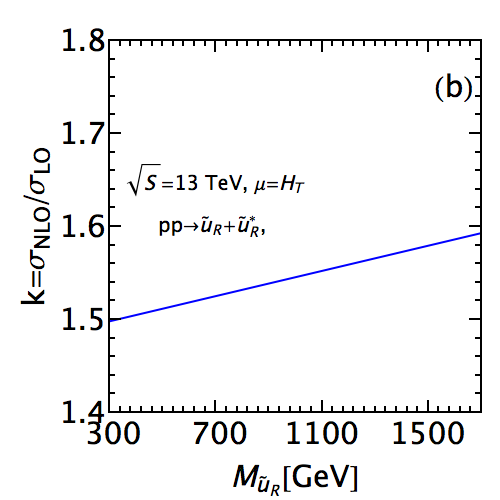}
	\caption{The production cross section (left) and $k$-factor (right) for mediator pair production via QCD at the 13 TeV LHC. Here, we have summed over contributions from the $ \tilde q \tilde q$, $ \tilde q \tilde q^*$ and $\tilde q^* \tilde q^*$ final states in each case.}
	\label{fig:csulul}
\end{figure}

The amplitude for pair production of mediators from initial state gluons is induced entirely from QCD interactions, whereas the quark-initiated subprocess receives both a QCD
contribution and one of order $g_{DM}^2$ from dark matter exchange.  The pure QCD contribution results in a $\tilde{q} \tilde{q}^*$ final state, and the only
unknown parameter entering into the rate is the mediator mass itself.  In Figure~\ref{fig:csulul}, we show the LO and NLO QCD cross sections 
at the $\sqrt{s} = 13$~TeV LHC as
a function of the mediator mass, as well as the $k$-factor, defined as $k \equiv \sigma_{NLO}/\sigma_{LO}$.  
At this order, the production cross section for the $q_L$ model is the sum of the results for the $u_R$ and $d_R$ models.
Evident from the figure, the $k$-factor is fairly flat in the mediator mass, with a value of about $\sim 1.5 ~{\rm to }~1.6$.

\begin{figure}\centering
	\includegraphics[width=0.475\linewidth]{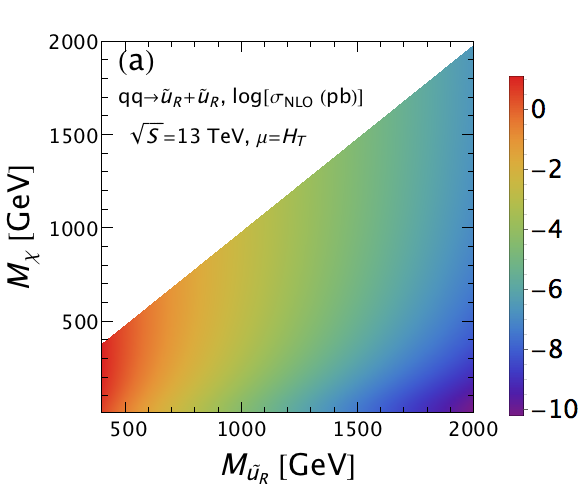}
	\includegraphics[width=0.475\linewidth]{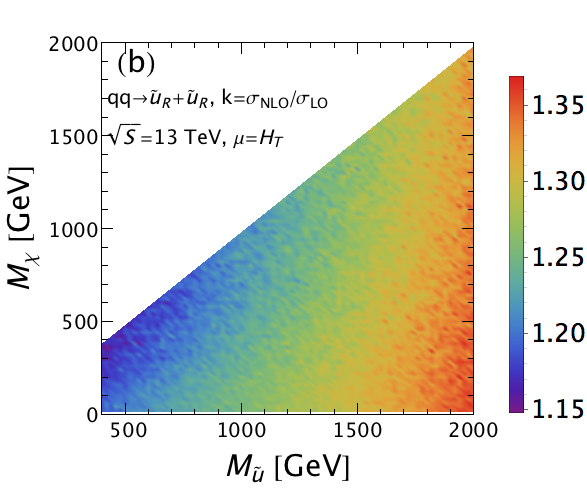}
	\includegraphics[width=0.475\linewidth]{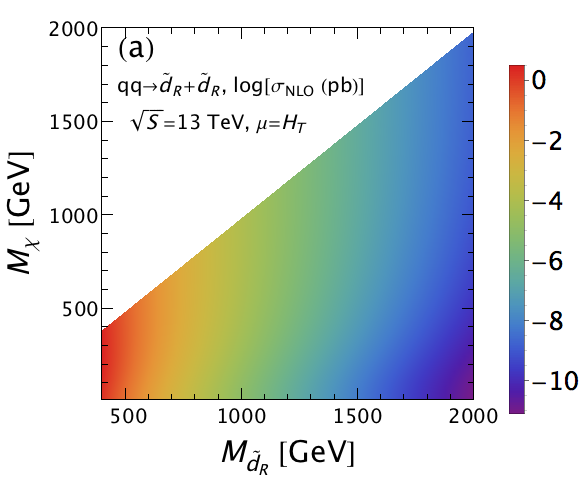}
	\includegraphics[width=0.475\linewidth]{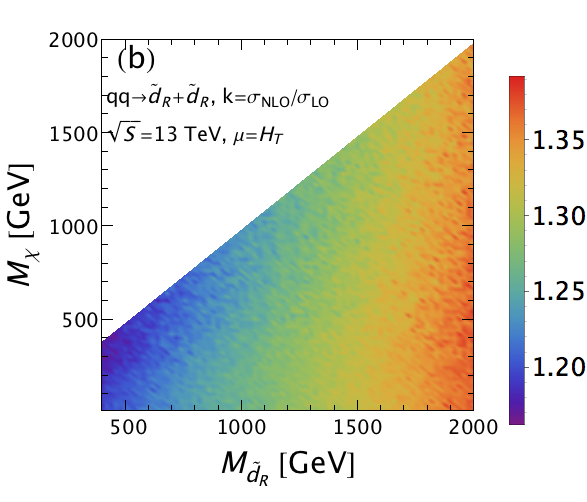}
	\caption{The production cross section  and $k$-factor for mediator pair production via dark matter exchange 
	in the $u_R$ model (upper) and $d_R$ model (lower) at the 13 TeV LHC.}
	\label{fig:csqqulul}
\end{figure}

The dark matter exchange contributions produce (for Majorana $\chi$) $\tilde{q} \tilde{q}^*$, $\tilde{q} \tilde{q}$, and $\tilde{q}^* \tilde{q}^*$ final states,
with rates which strongly depend on both $g_{DM}$ and $m_\chi$.  Since the scaling $\propto g_{DM}^4$ is simple, we set 
$g_{DM}=1$ as our benchmark point. 
The production cross section and $k$ factor in the $u_R$ and $d_R$ models are shown in the Fig.~\ref{fig:csqqulul}, summing over the 
$\tilde{q} \tilde{q}^*$, $\tilde{q} \tilde{q}$, and $\tilde{q}^* \tilde{q}^*$
final states in each case.
While the cross sections fall steeply as the mediator mass increases, they increase with larger $m_\chi$, as can be understood by the fact that the
$\tilde{q} \tilde{q}$ and $\tilde{q}^* \tilde{q}^*$ final states violate fermion number, and thus require an insertion of the Majorana mass to be nonzero.

\begin{figure}\centering
	\includegraphics[width=0.475\linewidth]{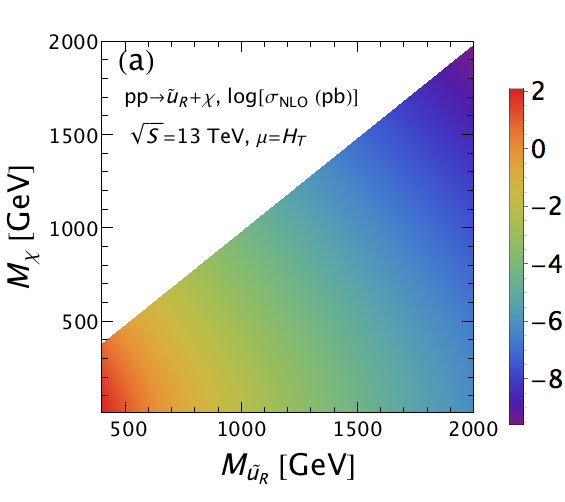}
	\includegraphics[width=0.475\linewidth]{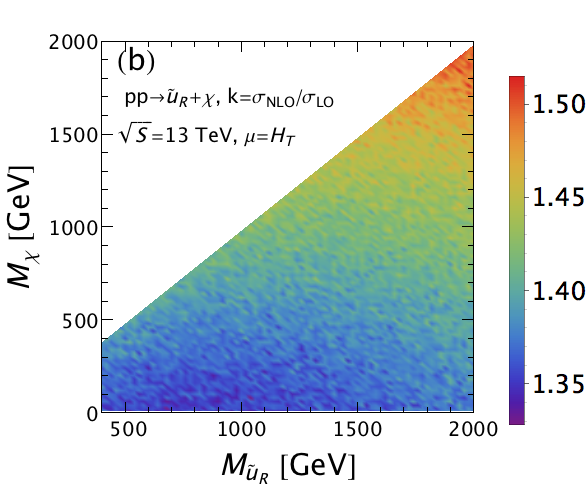}
	\includegraphics[width=0.475\linewidth]{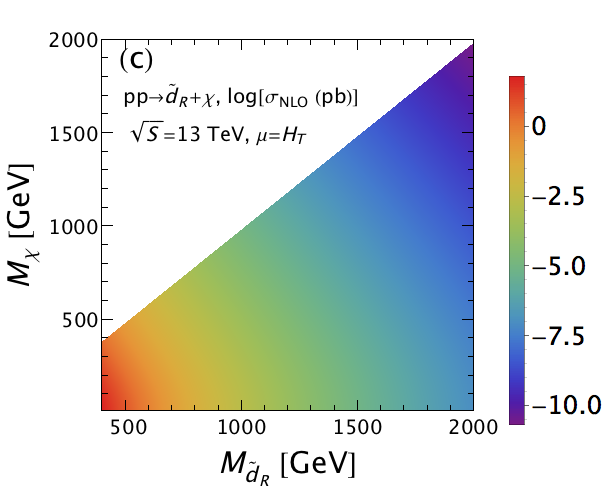}
	\includegraphics[width=0.475\linewidth]{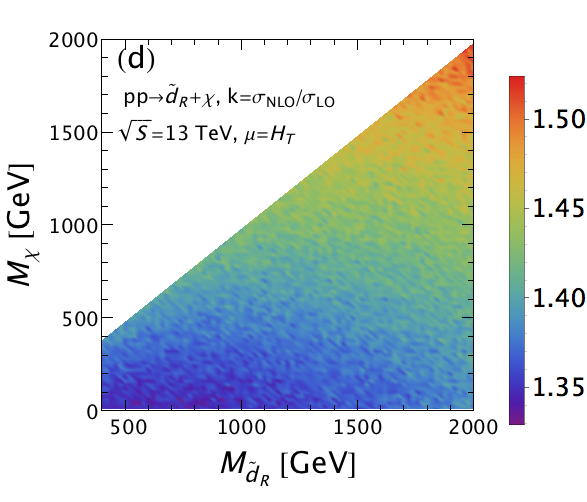}
	\caption{The production cross section  and $k$-factor for associated mediator plus dark matter production 
	in the $u_R$ model (upper) and $d_R$ model (lower) at the 13 TeV LHC.}
	\label{fig:csqnl}
\end{figure}

In Figure~\ref{fig:csqnl}, we show the NLO cross section for associated production of dark matter and a mediator in the $u_R$ and $d_R$  models. 
As the $s$-channel Feynman diagram dominates, the cross-section falls rapidly with increasing 
mediator plus dark matter mass.
However, we find that the $k$ factor increases with larger invariant mass from $\sim 1.35$ to $\sim 1.5$.

\subsection{Experimental Searches}

We assess the existing constraints from the null results of LHC searches for signatures including missing transverse momentum by implementing
two representative searches based on $\sim 36$~fb$^{-1}$ of integrated luminosity collected during operation at $\sqrt{s} = 13$~TeV:
\begin{itemize}
\item the mono-jet + $\slashed{E}_{T}$ search by ATLAS~\cite{Aaboud:2017phn}; and
\item the multi-jets + $\slashed{E}_{T}$ search by CMS~\cite{Sirunyan:2017cwe}.
\end{itemize}
In each case, we implement the search in the framework of {\tt Madanalysis5} \cite{Dumont:2014tja}, validate it 
against the experimental result, and re-cast limits on the simplified model parameter space.

\subsubsection{Mono-jet + $\slashed{E}_{T}$ search}

The ATLAS family of mono-jet + $\slashed{E}_{T}$ search~\cite{Aaboud:2017phn} selects events with:
\begin{itemize}
	\item a leading anti-$k_T$ jet with $p_{T}^{j}\ge 250$~GeV in the central region, $|\eta| < 2.4$;
	\item missing transverse momentum $\slashed{E}_{T} \geq 250$~GeV;
	\item up to four sub-leading anti-$k_T$ jets with $p_{T}> 30$~GeV and $|\eta| < 2.8$;
	\item $\Delta \phi ({\rm jet},\slashed{E}_{T})  > 0.4$ between the missing transverse momentum and any jet;
	\item no isolated electrons (muons) with  $p_T >20$ (10) GeV.
\end{itemize}
Events passing the selection cuts are further sorted into a variety of inclusive and exclusive bins in $\slashed{E}_{T}$.
Despite the moniker ``mono-jet", this search does not veto sub-leading jets, and is thus sensitive to $\slashed{E}_{T}$ plus one, two, or three jet topologies.

The analysis is  implemented within the \texttt{MadAnalysis5} framework, with the recasted code, as well as the details of validation are documented in the \texttt{MadAnalysis5} database \cite{recast:mono}.  
We follow the expirmental paper in order to recast the analysis.
  Anti-$k_T$ jets reconstructed  
using \texttt{FASTJET} \cite{Cacciari:2011ma}, with a jet radius of 0.4, and the detector simulated
by {\tt Delphes3}~\cite{deFavereau:2013fsa}, with parameters tuned to match the ATLAS specifications.
Since the detailed cut flows were not available for this analysis, it is validated by reproducing the exclusion curve for the supersymmetric process of scalar top production
with decay $\tilde{t}_{1}\to c \chi_{1}^{0} $\cite{Aaboud:2017phn}. 
The validated analysis is found to reproduce the experimental benchmark very closely.

Signal events for mediator pair and associated production are generated at tree level in \texttt{MadGraph5} with up to 
two additional jets, and subsequently showered and hadronized using \texttt{PYTHIA8} \cite{Sjostrand:2014zea}, with matrix element and parton shower (ME-PS) merging ~\cite{Hoche:2006ph}, 
performed using a merging parameter of $m_{\tilde{q}}/4$.  The rates are normalized to the NLO cross sections discussed above.
Based on the normalized number of events obtained for each signal region, we apply the log likelihood method to obtain an 
exclusion at each point in the $m_{\tilde{q}}-m_{\tilde{\chi}}$  plane.  \texttt{MadAnalysis5} is used to handle the event
selection and to compute the associated upper limit at the 95 $\%$ confidence level (CL) on
the signal cross section according to the CLs technique. 
Although the analysis contains a large number of signal regions, the upper bound on the cross section at each point
is determined predominantly from signal regions that have large signal rates, low background rates and small uncertainties \footnote{The bulk of the exclusion originates from signal regions IM8-IM10 and EM8-EM10, the high $p_{T}$ and $\slashed{E}_{T}$ regions  as described in \cite{Aaboud:2017phn}}.
\begin{figure}
	\centering
	\includegraphics[width=0.475\textwidth]{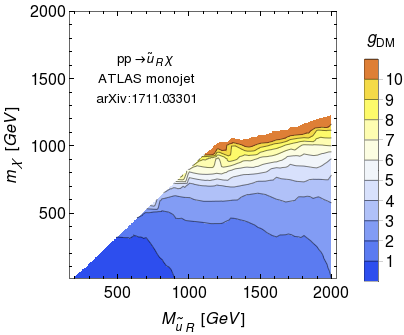}
	\includegraphics[width=0.475\textwidth]{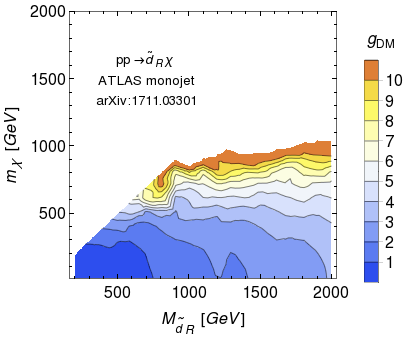}
	\includegraphics[width=0.475\textwidth]{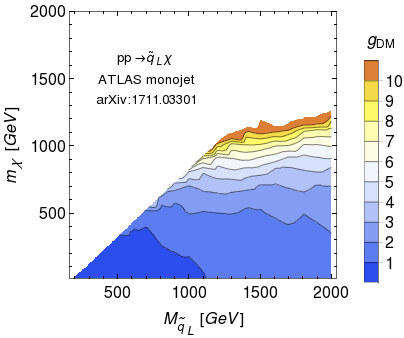}
	\caption{ATLAS mono-jet constraints on $g_{DM}$ derived from associated production for the $u_R$ (upper left), $d_R$(upper right) and $q_L$ (lower) models. Despite the moniker ``mono-jet'', this search does not veto
sub-leading jets, and is thus sensitive to $\slashed{E}_T$  plus one, two, or three jet topologies.
		\label{fig:monoMono}
	}
\end{figure}

\begin{figure}
	\centering
	\includegraphics[width=0.475\textwidth]{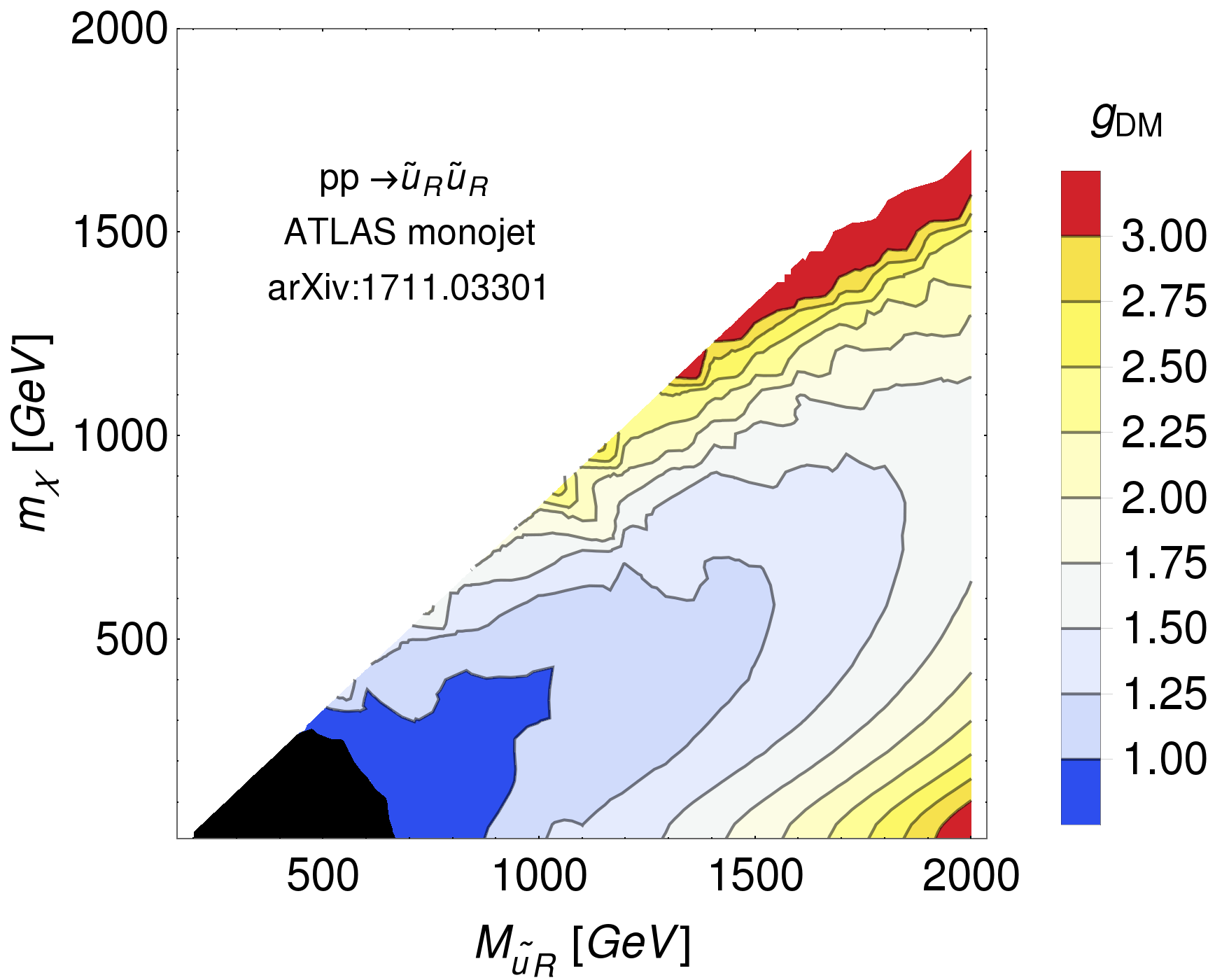}
	\includegraphics[width=0.475\textwidth]{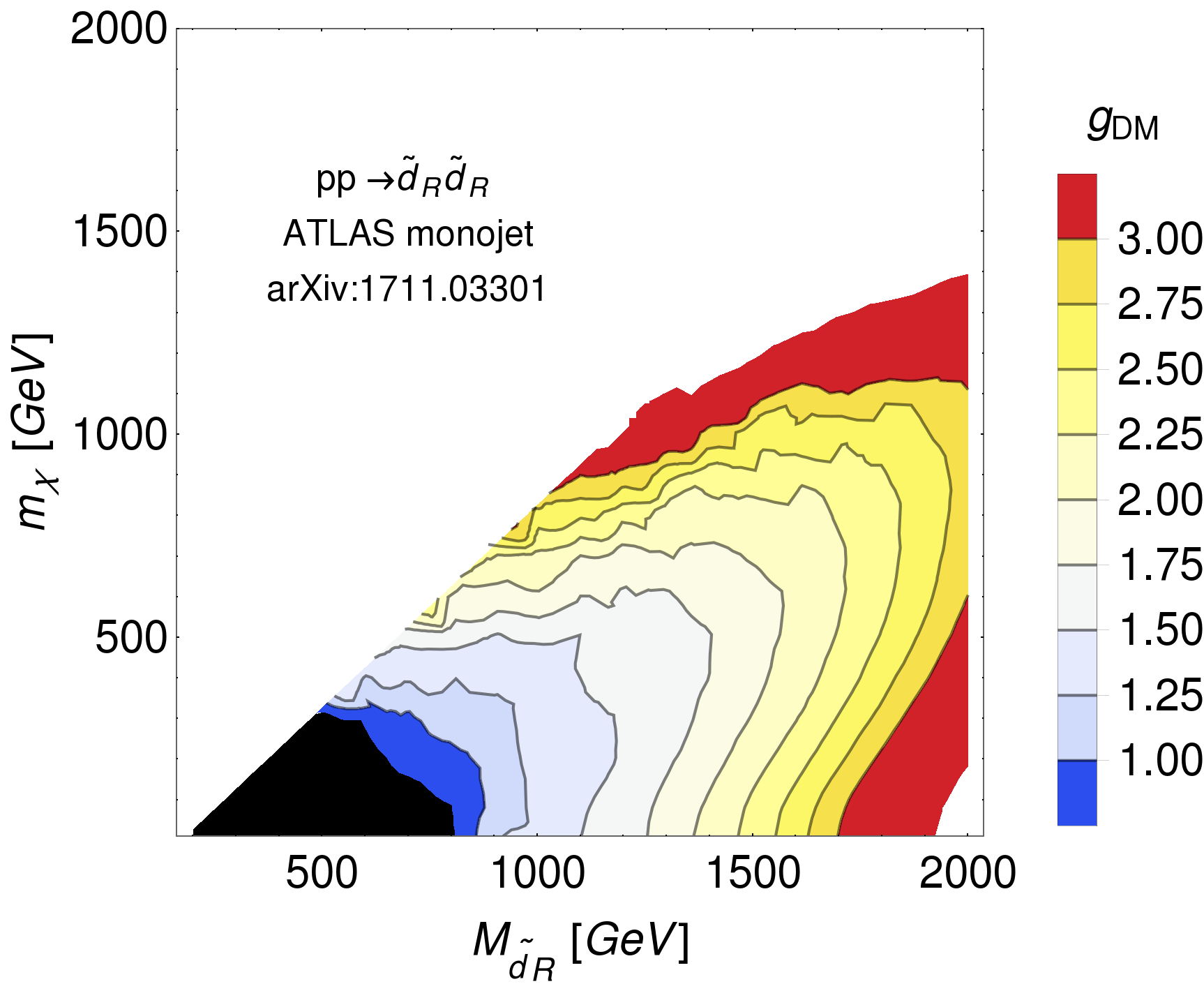}
	\includegraphics[width=0.475\textwidth]{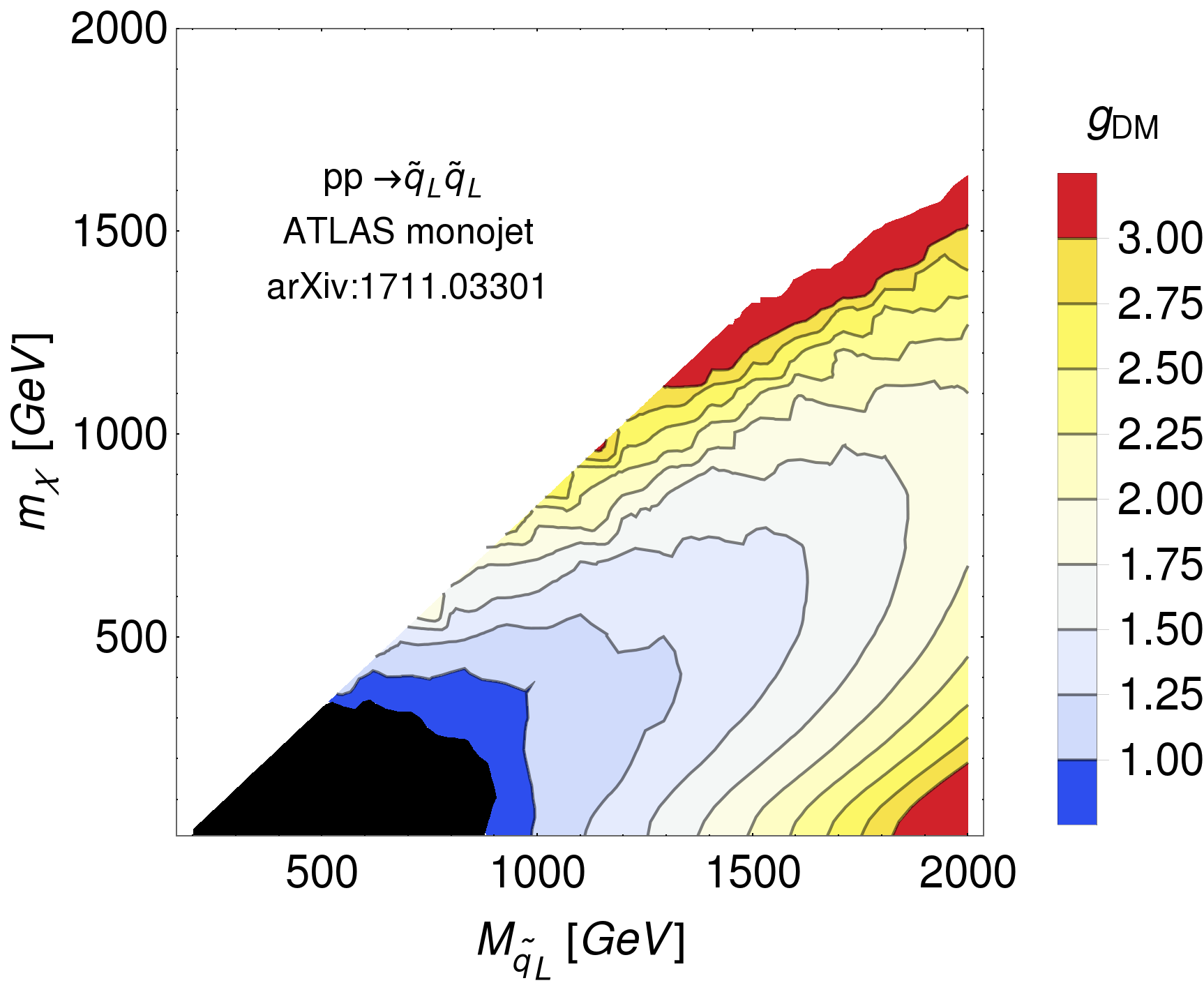}
	\includegraphics[width=0.41\textwidth]{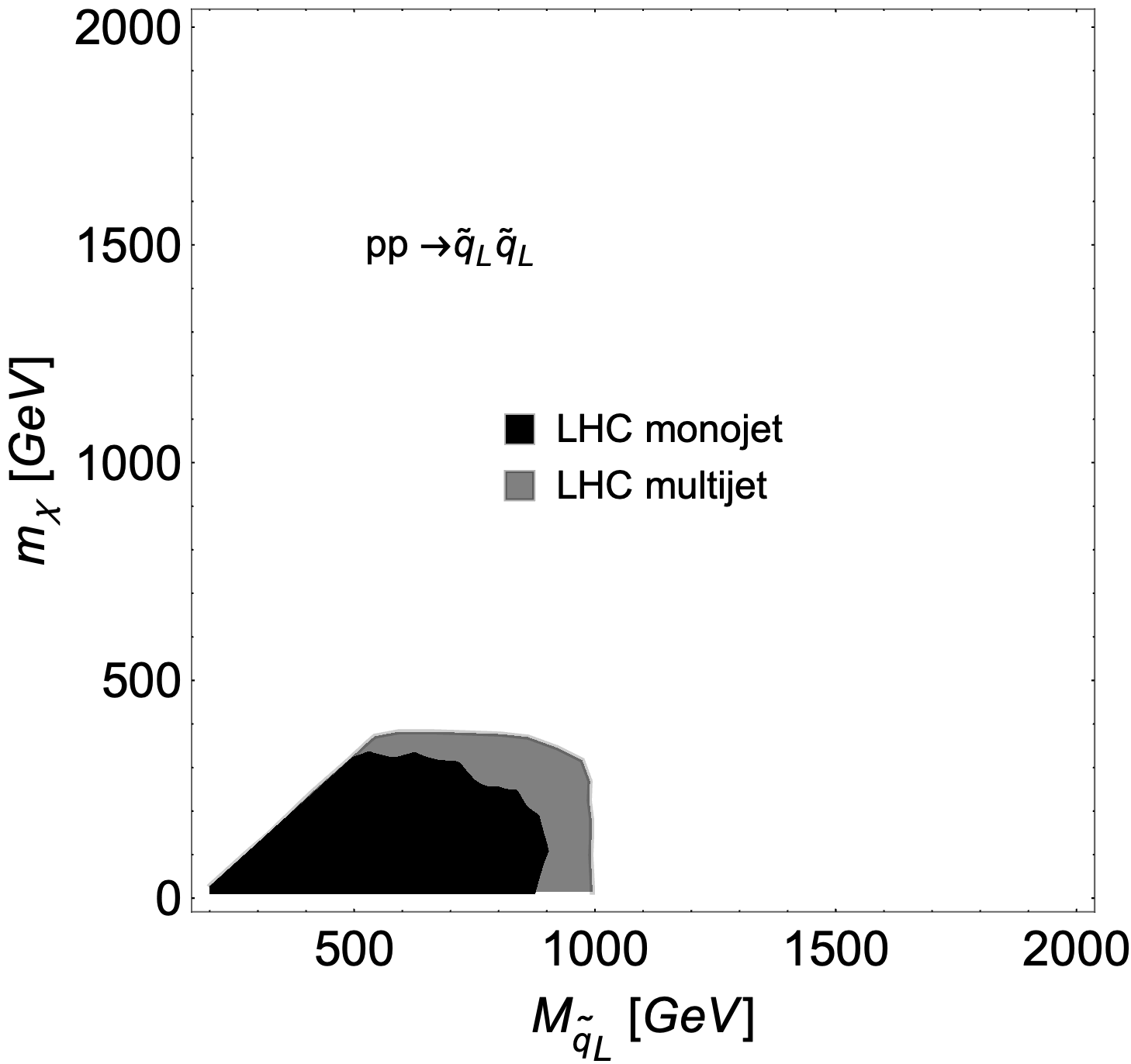}
	\caption{ATLAS monojet constraints on $g_{DM}$ derived from mediator pair production for the $u_R$ (upper left), $d_R$(upper right) and $q_L$ (lower left) models. Despite the moniker "mono-jet", this search does not veto
sub-leading jets, and is thus sensitive to $\slashed{E}_T$  plus one, two, or three jet topologies. In the lower right panel,
LHC monojet and multijet constraints for the $q_{L}$ model are compared. The black and grey regions 
	correspond to exclusion independent of the value of $g_{DM}$ for the two different experimental analyses, as indicated.
		\label{fig:pairMono}
	}
\end{figure}

In Fig.~\ref{fig:monoMono} we show the resulting constraints on the $u_R$, $d_R$ and $q_L$ simplified models
from analyzing the associated production process.  For most of the parameter space, the constraints are fairly weak, 
with the strongest constraints applying to the $q_L$ model.
Figure~\ref{fig:pairMono} shows constraints from mediator pair production.
While they are typically stronger than the associated production constraints, they are also rather weak for constraining most of the parameter space.  
However, for the lower range of masses considered, there is a region where QCD production of the mediators saturates the limit,
implying that these regions are decisively ruled out for any value of $g_{DM}$, 
resulting in a prompt mediator decay.
We also note that as one approaches the degenerate limit, $M_{\tilde{q}} \sim m_\chi$, the bounds from pair production rapidly
become ineffective as the jets from mediator decay become soft.
In addition, constraints get weaker for smaller $m_\chi$, which suppresses the $\tilde{q} \tilde{q}$ and $\tilde{q}^* \tilde{q}^*$ subprocesses
(this is particularly pronounced for larger mediator masses for which the $gg$ parton flux is less important relative to $qq$).  
In contrast, the associated production constraints of Fig.~\ref{fig:monoMono} has a much flatter behavior,
with constraints simply getting stronger with falling $m_\chi$.

\subsubsection{Multi-jet + $\slashed{E}_{T}$ search}

The  other LHC search of interest is the CMS search in the 0 lepton + jets + $\slashed{E}_{T}$ channel \cite{Sirunyan:2017cwe} with an 
integrated luminosity of 36.2 $\rm fb^{-1}$.  
It is designed to search for colored supersymmetric particles, gluinos and squarks, decaying as:
$\tilde{g}\to q\bar{q}\chi_{1}^{0}$, $\tilde{q}\to q \chi_{1}^{0}$ and $\tilde{t}_{1}\to t \chi_{1}^{0}$. 
A first set of baseline selections are applied as follows: $N_{j}\geq 2, ~H_{T} \geq 300~{\rm GeV}, ~{\rm MHT}> 300 ~{\rm GeV},~{\rm and}~\Delta(\phi,j) > 0.5$. 
Eventually, events are sorted into multiple signal regions in bins of the number of jets ($n_{j}$), 
the scalar sum of all jet transverse momenta ($H_{T}$), 
and the vector sum of all jet $p_T$ (MHT). For each bin of $n_{j}$, there are 10 different signal regions in bins of $H_{T}$ and MHT in intervals 
of 50 GeV, 150 GeV and 250 GeV. 
To facilitate validation,  in addition to the exclusion curves for specific channels, representative cutflows for a few benchmark points are provided. 
The analysis been recast within the framework of \texttt{MadAnalysis5} \cite{recast:multi}, where
it is found that the recast cutflows typically agree with the experimental ones to within 10$\%$.

Proceeding as before, constraints on the production of dark matter and/or mediators are derived from signal events produced
with \texttt{MadGraph5}.  The most important process is found to be mediator pair production, and constraints once again lead to
a region which is ruled out for any value of $g_{DM}$, resulting in a prompt mediator decay, because QCD production saturates the limits.
In the lower left panel of Fig.~\ref{fig:pairMono}, we compare the excluded region from the mono-jet and multi-jet searches.  The multi-jet search has constrained a somewhat
larger region in the plane of the mediator and dark matter masses. It should be noted that  one can also obtain the limits on $g_{DM}$
from this channel given that the limits are a bit more constraining than the monojet channel. However, since the direct detection constraints are 
significantly more constraining, we shall not perform that exercise here. 
\section{Combined Limits}
\label{sec:combination}

\begin{figure}
	\centering
	\includegraphics[width=0.475 \textwidth]{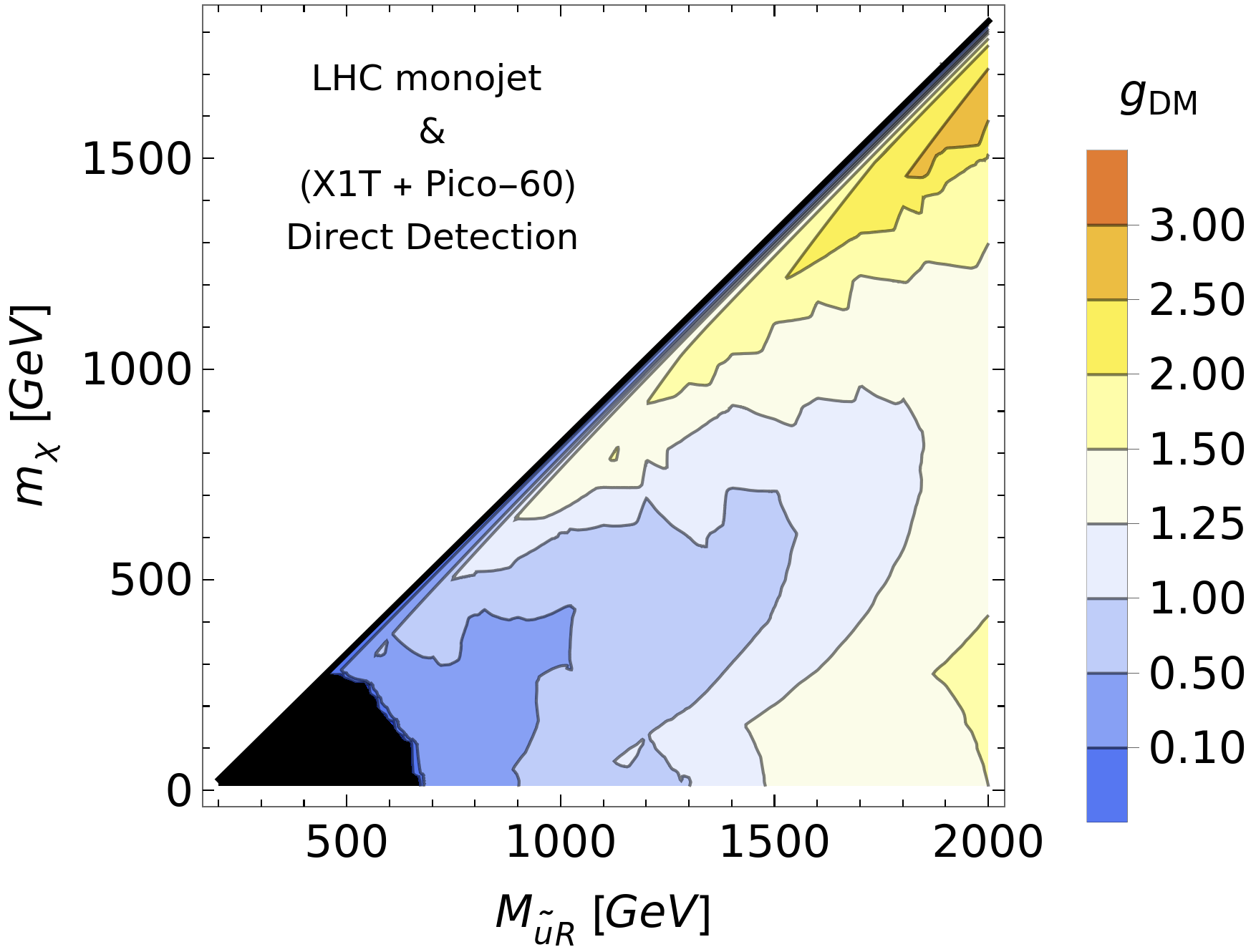}
	\includegraphics[width=0.475 \textwidth]{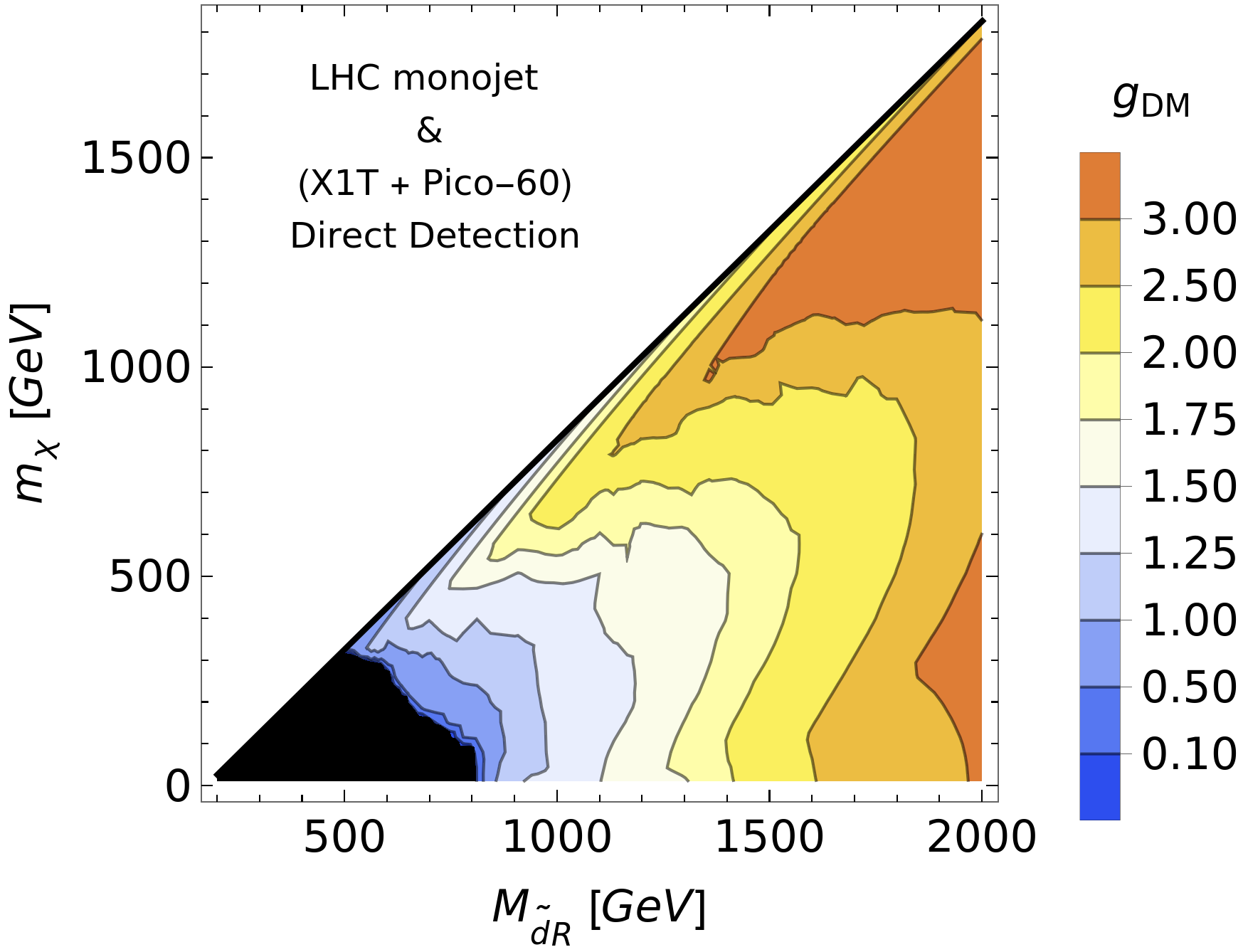}
	\includegraphics[width=0.475 \textwidth]{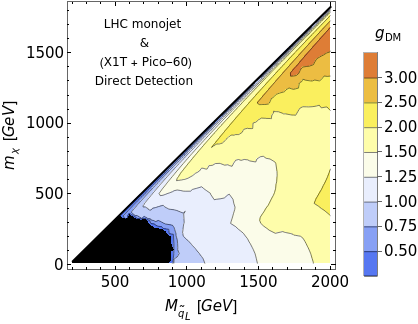}
	\caption{Combined SI, SD and  LHC constraints  for the $u_R$ (upper left), $d_R$(upper right) and $q_L$ (lower) models. 
	Shaded regions indicate allowed values of the coupling $g_{DM}$.
	\label{fig:combo}}
\end{figure}

We assemble the combined limits from direct detection and collider searches in Figure~\ref{fig:combo}.
The black shaded regions are ruled out for any value of $g_{DM}$ leading to prompt mediator decays, by constraints from the LHC, 
whereas colored shading indicates the maximum allowed $g_{DM}$ at that point.  
The picture that emerges is that collider and direct searches are highly complementary, with the collider able to rule out regions of parameter
space categorically, whereas the direct searches sensitive to SI scattering typically provide the strongest constraints on $g_{DM}$ in the region of parameter space where the mass of dark matter and mediator are comparable (with colliders filling the region of intermediate dark matter masses).  
Despite being less suppressed at tree level, the SD constraints are only relevant at large mediator mass ($\sim 2~{\rm TeV}$) and small dark matter mass ($\sim 10~{\rm GeV}$) and is subdominant for all other regions of parameter space. This highlights the importance that higher order
contributions to the SI cross section has on this particular simplified model.

\begin{figure}
	\centering
	\includegraphics[width=0.475 \textwidth]{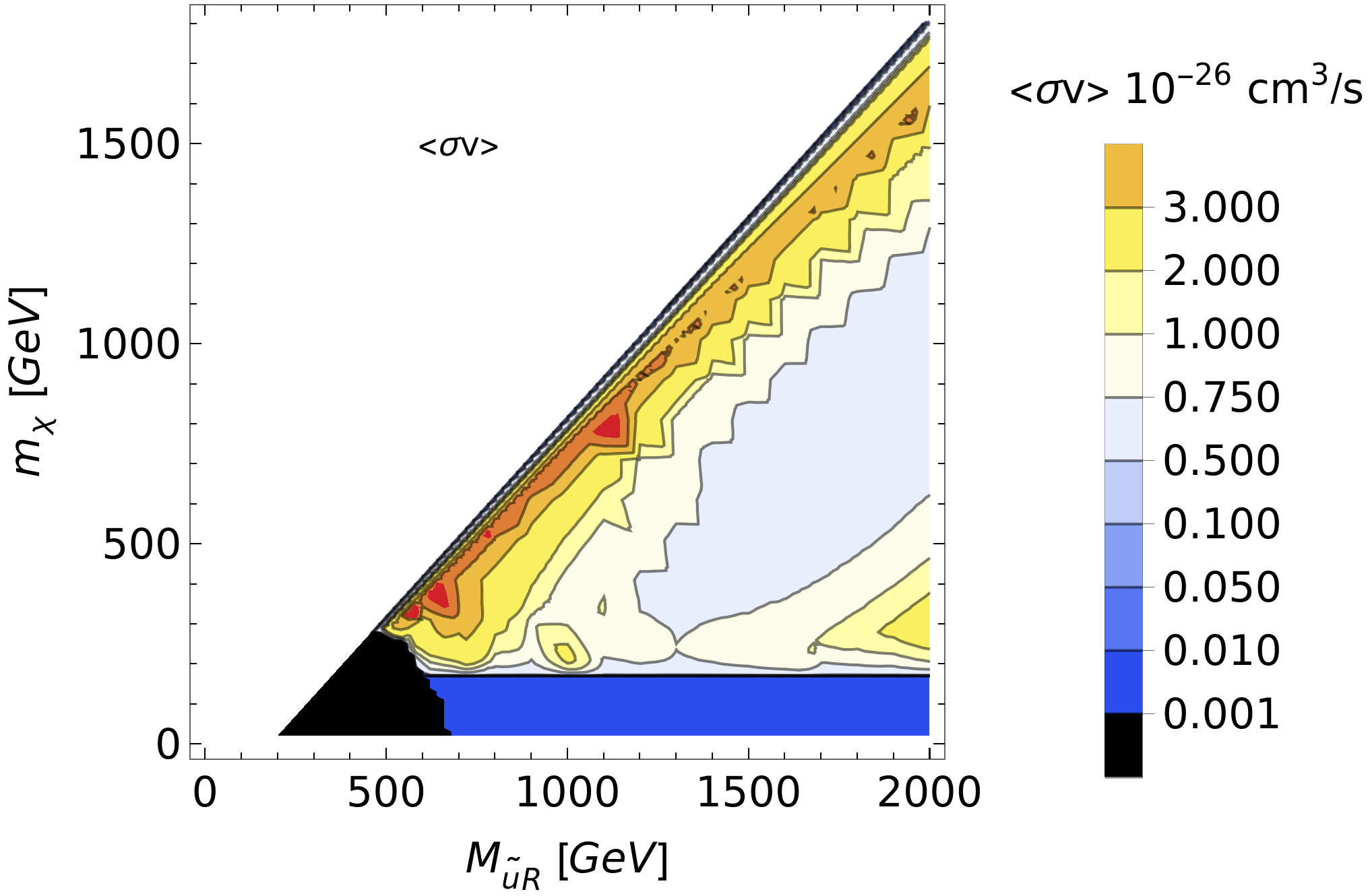}
	\includegraphics[width=0.475 \textwidth]{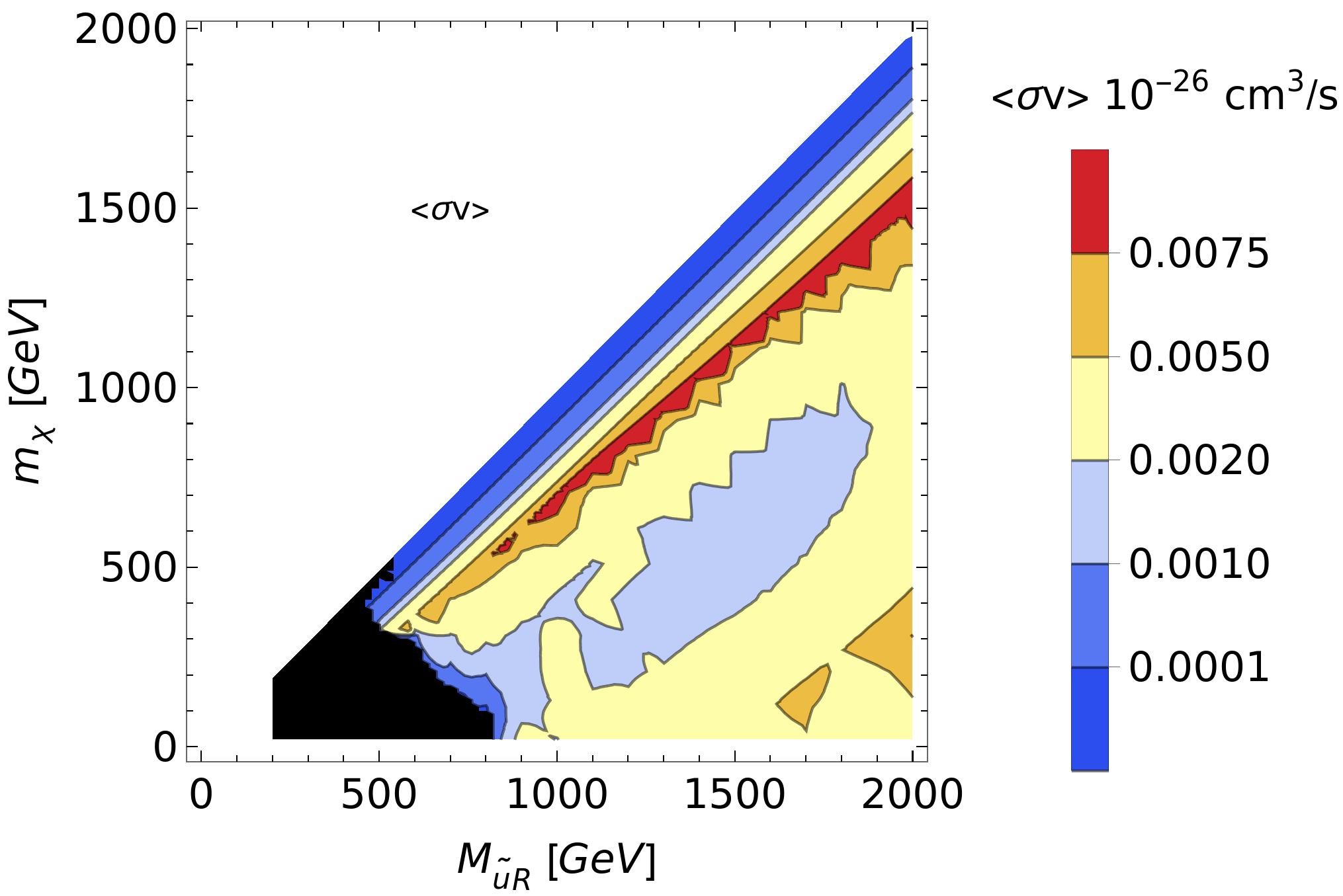}
	\includegraphics[width=0.475 \textwidth]{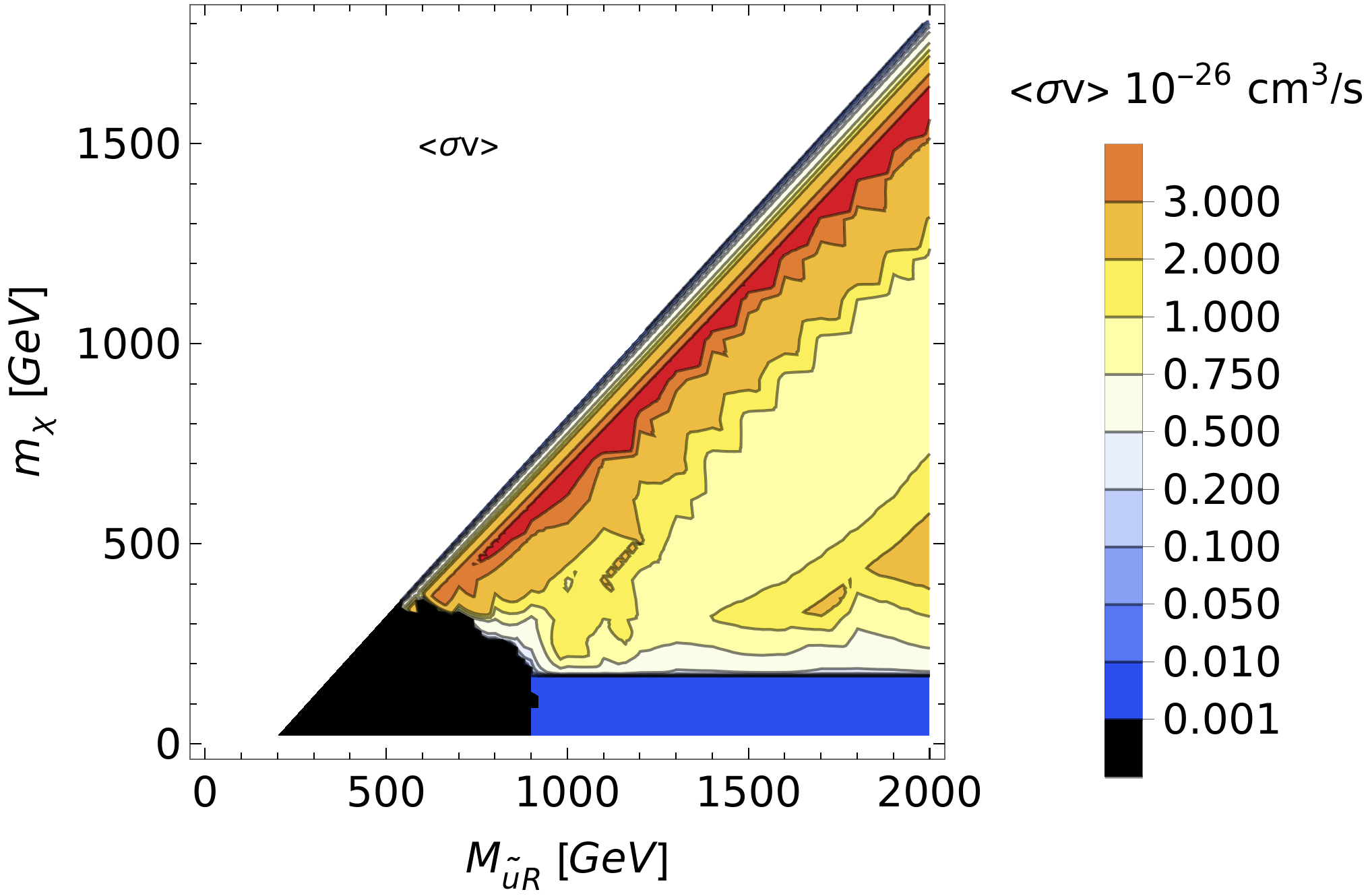}
	\caption{Velocity averaged annihilation cross-sections ($\left< \sigma v \right>$)  for the $u_R$ (upper left), $d_R$(upper right) and $q_L$ (lower) models in the zero temperature limit. 
		\label{fig:sigV}}
\end{figure}

Finally, it is interesting to use the current constraints from colliders and direct searches to construct the largest allowed forecast for the annihilation
cross section.  At very low velocities ($\beta \sim 10^{-3}$), this cross section is probed by indirect searches for high energy gamma rays, cosmic rays,
or neutrinos produced by dark matter annihilation in the Galaxy.  
In the early Universe, under the assumption that the extrapolation to the time of freeze-out follows a standard cosmology,
the inclusive annihilation cross section (for $\beta \sim 1/20$) maps onto the expected dark matter abundance.
In Figure~\ref{fig:sigV} we present the velocity averaged annihilation cross-section in the non relativistic limit. 
We import the model files written in Feynrules in {\tt micrOMEGAs5.0}~\cite{Belanger:2018mqt} 
to evaluate $\langle\sigma v_{ann}\rangle$ for $g_{DM}$ set to its maximally allowed value obtained from Figure~\ref{fig:combo}.
The black shaded area of Figure~\ref{fig:sigV} represents the region of parameter space ruled out by LHC constraints, and the
colored shaded regions correspond to different values of $\left< \sigma v \right>$ normalized to $10^{-26} \text{cm}^3/\text{s}.$\footnote{ A ballpark number for $\left< \sigma v \right>$ to saturate the DM relic density is  $3 \times 10^{-26} \text{cm}^3/\text{s}$, 
with smaller values indicating overabundant DM for a standard cosmological history.}  
Both the $q_L$ and $u_R$ models have larger values of $\left< \sigma v \right>$ compared to the $d_R$ model. This can be understood 
from the velocity-averaged annihilation cross-section into SM fermions, 
\begin{equation}
\left< \sigma v \right> \simeq N_{c}^{f} g_{DM}^{4} \Bigg[\frac{ m_{f}^{2} \sqrt{ 1- \frac{m_{f}^{2}}{m_{\chi}^{2}}} }{64\pi(m_{\tilde{q}}^{2} + m_{\chi}^{2}  - m_{f}^{2} )^{2}} + \beta^{2}\Bigg\{\frac{ m_{\chi}^{2}\sqrt{m_{\chi}^{4} + m_{\tilde{q}}^{4}} }{32\pi (m_{\chi}^{2} + m_{\tilde{q}}^{2})^{4} } + \mathcal{O}(m_{f}^{2})\Bigg\} \Bigg] \ ,
\end{equation}
where $N_{c}^{f}$ is the appropriate color factor for the species of fermion $f$, and $\beta$ is the velocity
of the colliding DM particles (Mandelstam $s =  4 m_\chi^2/(1  - \beta^2)$), which is  about $\sim 10^{-3}$. The first term is the velocity independent ($s$ wave scattering) part of the cross
section, while the second piece is the velocity dependent part of the annihilation ($p$ wave scattering).  For simplicity, in the term proportional to $\beta^2$, we only show the part of the expression that is independent of the quark mass ($m_f$).
The cross-section at zero velocity is proportional to the square of the quark mass, and in the $q_L$ and $u_R$ models is
dominated by annihilation into top quarks when kinematically accessible.
Annihilation to light quarks is dominated by the $p$-wave contribution which is proportional to $\beta^2$ and is therefore suppressed.
This is also the reason why $\langle\sigma v\rangle$ has a sharp increases for the $q_L$ and $u_R$ models at the top threshold where the $s$-wave dominates the contribution.
\section{Outlook}
\label{sec:outlook}

The identity of the dark matter remains one of the most pressing questions confronting particle physics, and the wealth of information from
colliders, searches for scattering with nuclei, and searches for dark matter annihilation complement each other in terms of making progress toward that
goal.  As the precision of the experimental searches increases, there is a need for a corresponding improvement in theoretical predictions,
in order to realize the full potential of the experimental data.

We have considered the class of $t$-channel models, in which the dark matter is a Majorana fermion which interacts with quarks at tree level
via exchange of colored scalar mediators.  At leading order, this model predicts no spin independent scattering with nuclei, suggesting that
searches for dark matter scattering with nuclei have much reduced power to constrain it.  However, at higher order there are contributions to
the spin independent cross section, which despite being suppressed, as compared to tree level SD cross-section, are far more constraining, especially in the degenerate region, because of the coherent enhancement of the experimental
limits compared to spin dependent searches.  We have assembled a consistent treatment of the Wilson coefficients leading to SI scattering, including
one loop contributions to the dark matter coupling to gluons and renormalization group resummation of large logs.  Our results highlight the
need for higher orders in analysis of the constraints on the parameter space of dark matter models.

Similarly, LHC searches provide key information which allows combinations of masses to be excluded for any value of the dark matter interaction
strength.  Understanding the reach of these searches accurately requires higher order calculations of the dark matter production processes.  We
have computed the rates for mediator pair and dark matter plus mediator associated production at next-to-leading order, and find that they
significantly improve the bounds extracted from LHC searches compared to the leading order results. While the efficiencies for the searches 
do not change, the evaluation of NLO cross sections improve the limits  by about 75-100~GeV. 

There are a few directions which are beyond the scope of this work, but would be worthwhile to explore.  We have largely neglected the
contributions from the top-flavored mediators, because their decays typically produce (on-shell or off-shell) top quarks, whose decays
themselves offer a wider range of final states, including high energy charged leptons.  There are also regions of parameter space
for which the mediator becomes long lived, with different (and powerful) constraints coming into play.  We look forward to incorporating
such refinements in the future. Finally, we would like to direct the interested reader, and in particular our experimental 
colleagues to the website of this project \cite{sdmwebsite}, where the analysis codes and additional plots can be found.

\section*{Acknowledgements}
The authors are grateful for conversations with Bakul Agarwal, R. Sekhar Chivukula, Tongyan Lin and Hua-sheng Shao. This work is supported in part by National Science Foundation
Grants No. PHY-1519045, PHY-1620638 and PHY-1719914. 
 C.-P. Yuan is also grateful for the support from the Wu-Ki Tung endowed chair in particle physics. 
This work was supported in
part by Michigan State University through computational resources provided by the Institute
for Cyber-Enabled Research

 \appendix

\section{Wilson Coefficients}
\label{app:WilsonCoefficients}
\subsection{Quarkionic Wilson Coefficients}
Representative Feynman diagrams for the LO contributions to the processes $q\chi \to q \chi$ are shown in  Figure~\ref{fig:feynqLO}, leading to amplitude
\begin{gather}
\mathcal{M} = (-ig_{\!_{DM}})^2 (\bar{\chi}P_Ru) \frac{i}{p^2-M_{\tilde{u}}^2} (\bar{u}P_L\chi) \nonumber\\
\approx \left(\frac{ig_{\!_{DM}}^2}{M^2_{\tilde{u}}-m^2_{\chi}} + 2 \frac{k_{q\nu} k_{\chi}^{\nu}}{(M^2_{\tilde{u}}-m^2_{\chi})^2}\right) \frac{1}{8} [(\bar{\chi}\gamma^{\mu}\chi)(\bar{u}\gamma_{\mu}u) - (\bar{\chi}\gamma^{\mu}\gamma^5\chi)(\bar{u}\gamma_{\mu}\gamma_5u)] .
\end{gather}
Here $k_q$ and $k_\chi$ are the four momenta of the quark and dark matter respectively.
Terms that are order $(M^2_{\tilde{u}}-m^2_{\chi})^2$ in the expansion above have explicit momentum factors in the numerator. This matches to the matrix elements generated from the effective lagrangian in Equation~(\ref{eq:lag-eft}) from which we determine the wilson coefficients given in Equation~(\ref{eq:treesq}).

\subsection{Gluonic Wilson Coefficients}
\label{sec:gulonic-full}

Representative Feynman diagrams for the LO contributions to the process $g\chi \to g \chi$ are shown in Figure~\ref{fig:Feynman-gNLO}.
Amplitudes are generated with {\tt Feynarts} \cite{Hahn:2000kx}, decomposed into tensor integrals 
using {\tt FormCalc} \cite{Hahn:1998yk}, and the expansion of tensor integrals is performed with {\tt Package-X} \cite{Patel:2015tea}. In order to calculate Wilson coefficients we match amplitudes produced from the full theory in Equations~(\ref{eq:ur}-\ref{eq:ql}) to that from the EFT of Equation~(\ref{eq:lag-eft}). We calculate $S$-channel production $g(k_1) g(k_2) \to \chi(k_3)\chi(k_4)$ and apply the following projection operators to the amplitude to in order to match amplitudes:
\begin{equation}
A^{\mu_1\mu_2} = i 8f_{G} \left(k_2{}^{\mu _1} k_1{}^{\mu _2}-g^{\mu _1\mu _2} \left(k_1\cdot k_2\right)\right) \ ,
\end{equation}
\begin{eqnarray}
B^{\mu_1\mu_2} &=&i \frac{g^{(2)}_{G}}{m_\chi^2} 
\bigg[
g^{\mu _1\mu _2} \bigg(2 (k_1\cdot k_3) (k_2\cdot k_3)+2 (k_1\cdot k_4) (k_2\cdot k_4)-(k_3^2+k_4^2) (k_1\cdot k_2)\bigg)\nn
&+& k_1^{\mu _2} \bigg((k_3^2+k_4^2) k_2^{\mu _1}-2 (k_3^{\mu _1} (k_2\cdot k_3) 
+k_4^{\mu _1} (k_2\cdot k_4))\bigg)\nn
&+&k_3^{\mu _2} \bigg(2 k_3^{\mu _1} (k_1\cdot k_2)-2 k_2^{\mu _1} (k_1\cdot k_3)\bigg) \nn
&+&2 k_4^{\mu _2} \bigg(k_4^{\mu _1} (k_1\cdot k_2)-k_2^{\mu _1} (k_1\cdot k_4)\bigg)\bigg]\ ,
\end{eqnarray}
and,
\begin{eqnarray}
C^{\mu_1\mu_2}&=& i 2\frac{g^{(1)}_{G}}{m_\chi}  \bigg[
g^{\mu _1\mu _2} (k_1\cdot k_3) \gamma \cdot k_2-g^{\mu _1\mu _2} (k_1\cdot k_4) \gamma \cdot k_2+g^{\mu _1\mu _2} (k_2\cdot k_3) \gamma \cdot k_1 \nn
&-&g^{\mu _1\mu _2} (k_2\cdot k_4) \gamma \cdot k_1+(k_1\cdot k_2) (g^{\mu _1\mu _2} (\gamma \cdot k_4-\gamma \cdot k_3) 
+\gamma ^{\mu _2} (k_3^{\mu _1}-k_4^{\mu _1})+\gamma ^{\mu _1} (k_3^{\mu _2}-k_4^{\mu _2})) \nn
&-&\gamma ^{\mu _2} k_2^{\mu _1} (k_1\cdot k_3)+\gamma ^{\mu _2} k_2^{\mu _1} (k_1\cdot k_4) - k_3^{\mu _2} k_2^{\mu _1} (\gamma \cdot k_1)+k_4^{\mu _2} k_2^{\mu _1} \gamma \cdot k_1 \nn
&+&k_1^{\mu _2} (\gamma ^{\mu _1} (k_2\cdot k_4-k_2\cdot k_3) 
+k_2^{\mu _1} (\gamma \cdot k_3-\gamma \cdot k_4)+(k_4^{\mu _1}-k_3^{\mu _1}) \gamma \cdot k_2) 
\bigg]\ .
\end{eqnarray}
The three projection operators correspond to the 
three $gg\chi\chi$ vertices orignating from the operators in the effective Lagrangian in Equation~(\ref{eq:lag-eft}). 

We define Mandelstam variables as usual:
\begin{eqnarray}
&S&=(k_1+k_2)^2,\quad T=(k_1-k_3)^2, \quad U=(k_1-k_4)^2 ,
\end{eqnarray}
which satisfy~$S+T+U = 2 m_{\chi}^2$.
To simplify the matching we choose a specific phase space point --  $U=m_\chi^2 -S/2$ and $T= m_{\chi}^2 -S/2$.
The resulting expressions are expanded in powers of $S$ and we extract the Wilson coefficients by matching to the following equations.
	\begin{eqnarray}
	A\cdot (A + B + C) &=& 32 f_{G}^2 S^2\nn
	B\cdot (A + B + C) &=& -2 m_\chi^3 S^2(m_\chi \frac{g^{(2)}_{G}}{m_\chi^2}  + 2 \frac{g^{(1)}_{G}}{m_\chi} ) + \mathcal{O}(S^3)\nn 
	C\cdot (A +  B +C) &=& 2 m_\chi^2 S^2(m_\chi \frac{g^{(2)}_{G}}{m_\chi^2}  + 4 \frac{g^{(1)}_{G}}{m_\chi} )  + \mathcal{O}(S^3)
	\end{eqnarray}

The Wilson coefficents for the gluonic operators listed in Equation~(\ref{eq:lag-eft}) are found to be, 
\begin{eqnarray}
f_{G}&=&\alpha_s g_{DM}^2
m_{\chi}\bigg[ -12 m^2 M^4 m_{\chi}^2
(m^2-M^2+m_{\chi}^2) \Lambda
(m_{\chi}^2;m,M)\nn
&-&(m-M-m_{\chi}) (m+M-m_{\chi})
(m-M+m_{\chi}) (m+M+m_{\chi}) \nn
&\times&
\big\{m^6-3 m^4 (2
M^2+m_{\chi}^2)+m^2 (3 M^4+2 M^2 m_{\chi}^2+3
m_{\chi}^4)+(M^2-m_{\chi}^2)^2 
(2
M^2-m_{\chi}^2)\big\}\bigg]\nn
&\times&\frac{1}{192 \pi  M^2 (m-M-m_{\chi})^3
	(m+M-m_{\chi})^3 (m-M+m_{\chi})^3 (m+M+m_{\chi})^3}\ ,
\label{eq:fGfull}
\end{eqnarray}
where
\begin{eqnarray}
\Lambda(m_{\chi}^2;m,M)&= &
\frac{\lambda}{m_{\chi}^2} \log \left(\frac{m^2+\lambda+M^2-m_{\chi}^2}{2 m
	M}\right)\nn
\lambda&=& \sqrt{m^4-2 m^2 M^2-2 m^2 m_{\chi}^2+M^4-2 M^2
	m_{\chi}^2+m_{\chi}^4} \ .
\end{eqnarray}	
Also,
\begin{eqnarray}
\frac{g^{(2)}_{G}}{m_\chi^2}&=&\alpha_s g_{DM}^2
\bigg[2 m_{\chi}^2 \big\{10 m_{\chi}^4
(m^6-M^6)+m_{\chi}^8 (m^2-5 M^2)
-5 m_{\chi}^2 (m^2-M^2)^3 (m^2+M^2)\nn
&+&(m^2-M^2)^5 + 2 m_{\chi}^6 (-4
m^4+2 m^2 M^2+5 M^4)+m_{\chi}^{10}\big\} \Lambda
(m_{\chi}^2;m,M)\nn
&-&(m-M-m_{\chi}) (m+M-m_{\chi})
(m-M+m_{\chi}) (m+M+m_{\chi}) \nn
&\times&\bigg\{7 m_{\chi}^4
(m^4-M^4)-2 m_{\chi}^6 (m^2-4 M^2)-2
m_{\chi}^2 (m^2-M^2)^3\nn
&+&2 (m^4-2 m^2
(M^2+m_{\chi}^2)+(M^2-m_{\chi}^2)^2)^2
\log (\frac{m}{M})-3 m_{\chi}^8\bigg\}\bigg]\nn
&\times&\frac{1}{48 \pi 
	m_{\chi}^5 (m-M-m_{\chi})^3 (m+M-m_{\chi})^3
	(m-M+m_{\chi})^3 (m+M+m_{\chi})^3} ,~
\label{eq:gG2full}
\end{eqnarray}
\begin{eqnarray}
\frac{g^{(1)}_{G}}{m_\chi}&=&\alpha_s g_{DM}^2
\bigg[2 m_{\chi}^2 (3 m_{\chi}^2
(M^4-m^4)+m_{\chi}^4 (5
m^2+M^2)+(m^2-M^2)^3-3 m_{\chi}^6) \Lambda
(m_{\chi}^2;m,M)\nn
&+&2 (m+M-m_{\chi}) (m-M+m_{\chi})
(m+M+m_{\chi}) \bigg\{m_{\chi}^2 (m-M-m_{\chi}) (m^2-M^2-3
m_{\chi}^2)\nn
&-&(m+M-m_{\chi}) (m-M+m_{\chi})
(-m+M+m_{\chi})^2 (m+M+m_{\chi}) \log
(\frac{m}{M})\bigg\}\bigg]\nn
&\times&	\frac{1}{192 \pi  m_{\chi}^4
	(m+M-m_{\chi})^2 (m-M+m_{\chi})^2 (-m+M+m_{\chi})^2
	(m+M+m_{\chi})^2} .
\label{eq:gG1full}
\end{eqnarray}

\subsection{Gluonic Wilson Coefficients using the Fock-Schwinger Gauge}

The determination of gluonic Wilson Coefficients can be greatly simplified by using the Fock-Schwinger gauge (for details, see for e.g., References~\cite{Novikov:1983gd} and \cite{Hisano:2010ct} and references therein),
for which 
\begin{equation}
x^\mu A_{\mu}^{a}(x)= 0\ .
\end{equation}
In this gauge one can express the gluon field in terms of its field strength tensor $G^{\mu\nu}$, maintaining explicit gauge invariance at each step.
 The Wilson coefficients can be extracted from the one loop contribution to a Majorana fermion propagating through a
 non-zero background of gluon fields. The gluonic background modifies the form of the internal quark and mediator propagators.
Calculations are performed using {\tt FeynCalc-9.20}~\cite{Shtabovenko:2016sxi}, and we summarize the results
 corresponding to a single flavor of quark here.

 \begin{equation}
f_G=
\frac{\alpha_s}{4\pi}\frac{1}{8} \left( 
\int \frac{d^4q}{i \pi^2}
\frac{\slashed{p} + \slashed{q}}{((p+q)^2-m_q^2)^4(q^2-M_{\tilde{q}}^2)} +
\int \frac{d^4q}{i \pi^2} 
\frac{\slashed{p} + \slashed{q}}{((p+q)^2-m_q^2)(q^2-M_{\tilde{q}}^2)^4} \right)  \ .
\label{fsdfld}
\end{equation}
These integrals maybe solved by introducing Feynman parameters to yield the result in Equation~(\ref{eq:fGfull}).

For the twist-2 operators we can write out the scattering amplitude as follows and then use projection operators to match and find the Wilson coefficients.
 \begin{eqnarray}
 \mathcal{M}_{t2}=
 &-&\frac{\alpha_s}{4\pi} \bar{\chi}
 \int \frac{d^4q}{i \pi^2}\frac{1}{8} 
 \frac{(\slashed{p} + \slashed{q})((3m_{\tilde{q}}^2- q^2) g^{\mu\nu} - 4 q^{\mu}q^{\nu})}{((p+q)^2-m_q^2)^4(q^2-m_{\tilde{q}}^2)}\chi G_{\mu}^{a\rho}G_{\rho\nu}^{a} 
 \nonumber\\
 &-&\frac{\alpha_s}{4\pi} \bar{\chi}
 \int \frac{d^4q}{i \pi^2}\frac{1}{4} 
 \frac{(\slashed{p} + \slashed{q})((m_{q}^2-2(p+q)^{\mu}(p+q)^{\nu}) }{((p+q)^2-m_q^2)(q^2-m_{\tilde{q}}^2)^4}\chi G_{\mu}^{a\rho}G_{\rho\nu}^{a}
 \nonumber\\
 &+&\frac{\alpha_s}{4\pi} \bar{\chi}
 \int \frac{d^4q}{i \pi^2}\frac{1}{4} 
 \frac{(\gamma^{\nu}(p+q)^{\mu} + \gamma^{\mu}(p+q)^{\nu})}{((p+q)^2-m_q^2)(q^2-m_{\tilde{q}}^2)^3}\chi G_{\mu}^{a\rho}G_{\rho\nu}^{a} ,
 \end{eqnarray}
The integrals above can be solved by introducting Feynman parameters and the final result is given in Equations~(\ref{eq:gG2full}) and (\ref{eq:gG1full}).

\section{Numerical Values}
\label{app:parameters}
We list here the various values that we have used in our numerical analysis. Light quark masses are taken from PDG~\cite{Tanabashi:2018oca} and are defined in the $\overline{\rm MS}$ scheme at $\mu=2~{\rm GeV}$.
\begin{align}
&
m_u = 2.2 ~{\rm MeV},\quad 
m_d = 4.7~ {\rm MeV},\quad
m_s = 95~ {\rm MeV},\ 
& \nonumber\\
&
m_c = 1.3~ {\rm GeV},\quad 
m_b = 4.2~{\rm GeV},\quad 
m_t = 172~{\rm GeV},
& \nonumber\\
&
m_Z = 91.188~{\rm GeV},\quad 
\alpha_s(m_Z) = 0.1184,\quad 
&\nonumber \\
&
m_n = 0.9396~{\rm GeV}\quad 
m_p = 0.9383~{\rm GeV}\ .&
\end{align}

Values of hadronic matrix elements for the spin-0 operators evaluated at the scale $\mu=2~{\rm GeV}$ are taken from reference~\cite{Hill:2014yxa} and are given below \footnote{There are recent calculations for these parameters, see also \cite{Alarcon:2011zs,Alarcon:2012nr} }. 
\begin{align}
&
\left[f_{T_u}\right]_p = 0.018,\quad 
\left[f_{T_d}\right]_p = 0.030,\quad 
\left[f_{T_s}\right]_p = 0.043, 
&\nonumber \\
&
\left[f_{T_u}\right]_n = 0.015,\quad 
\left[f_{T_d}\right]_n = 0.034,\quad 
\left[f_{T_s}\right]_n = 0.043, 
&\nonumber \\
&
f_{T_G}|_{\rm NNNLO} = 0.80\ .
&
\end{align}
Here $\left[f_{T_x}\right]_y$ corresponds to the contribution of the $x$ quark to the  nucleon matrix elements for the nucleon $y$.  $f_{T_G}$ is determined using the sum rule
\begin{equation}
f_{T_G} = -\frac{9 \alpha_{S}(\mu)}{4 \pi \beta(\mu)}\left[
1 - \left( 1 + \gamma_m(\mu)\right)\sum_{u,d,s}f_{Tq} 
\right]\ .
\end{equation} 
Here $\beta(\mu)$ and $\gamma_m (\mu)$ are the QCD beta function and quark anomalous dimension respectively.
 Here we calculate $f_{T_G}$ by using expressions of $\beta(\mu)$ and $\gamma_m(\mu)$ up to order  ${\rm N}^3{\rm LO}$ in $\alpha_s$. For details on how to calculate $f_{T_G}$, see for example, Reference~\cite{Hill:2014yxa}.
Hadronic matrix elements for twist-2 operators defined in Equation~(\ref{eq:lag-eft}), also defined at the scale
$\mu = 2$ GeV, are extracted from the CT14NNLO parton distribution functions~\cite{Dulat:2015mca}.
\begin{align}
&
\left[u(2) + \bar{u}(2)\right]_p = 0.3481,\quad
\left[d(2) + \bar{d}(2)\right]_p = 0.1902,\quad
&\nonumber \\
&
\left[s(2) + \bar{s}(2)\right]_p = 0.0352,\quad 
\left[c(2) + \bar{c}(2)\right]_p = 0.0107\ ,
&\nonumber \\
&
\left[G(2)\right]_p=\left[G(2)\right]_n=0.4159\ .
&
\end{align}
When evaluating spin dependent cross-sections we use the following parameters for nuclear axial vector currents~\cite{Freytsis:2010ne,Nakamura:2010zzi}
\begin{align}
&
\Delta u^{(p)}=0.84,\quad
\Delta d^{(p)}=-0.43,\quad
\Delta s^{(p)}= -0.09 ,
&\nonumber \\
&
\Delta u^{(n)}= \Delta d^{(p)},\quad
\Delta d^{(n)}= \Delta u^{(p)},\quad
\Delta s^{(n)}=  \Delta s^{(p)}.\ 
& 
\end{align}

We approximate all of isotopes of Xenon, in the  Xenon 1T experiment, to have a mass number and atomic number
$A_{Xe} = 131$ and $Z_{Xe} = 54$ respectively.
\bibliographystyle{utphys}
\bibliography{ref.bib}

\end{document}